\documentclass[useAMS,usenatbib,usegraphicx]{mn2e}
\usepackage{times}
\usepackage{lscape}
\usepackage{amssymb}
\def\kms{$\rm km\, s^{-1}$}
\def\cm3{$\rm cm^{-3}$}

\def\n0{$\rm n_{0}$}
\def\B0{$\rm B_{0}$}

\def\mc{$\mu$m}

\def\L12{L$_{12\mu m}$~}
\def\F12{F$_{12\mu m}$~}

\def\fe2{[Fe\,{\sc ii}]}
\def\s3{[S{\sc iii}]}
\def\h2{H$_{2}$}
\def\w{$W_{\lambda}$}
\def\F{$F_{\lambda}$}

\def\pp{$\pm$}

\DeclareRobustCommand{\ion}[2]{%
\relax\ifmmode
\ifx\testbx\f@series
{\mathbf{#1\,\mathsc{#2}}}\else
{\mathrm{#1\,\mathsc{#2}}}\fi
\else\textup{#1\,{\mdseries\textsc{#2}}}%
\fi}

 \title[NIR integrated spectra of Galactic GCs]{Near-infrared integrated spectra of Galactic globular clusters: testing simple stellar population models}

\author[Riffel et al.]{R. Riffel$^{1}$\thanks{E-mail:
riffel@ufrgs.br}, D. Ruschel-Dutra$^{1}$, M. G. Pastoriza$^{1}$, A. Rodr\'{i}guez-Ardila$^{2}$,
\newauthor J. F. C. Santos Jr.$^{3}$, C. J. Bonatto$^{1}$ and J. R. Ducati$^{1}$
\\$^{1}$Departamento de Astronomia, Universidade Federal do Rio Grande do Sul. Av. Bento Gon\c calves 9500, Porto Alegre, RS, Brazil.
\\$^2$ Laborat\'{o}rio Nacional de Astrof\'{i}sica/MCT - Rua dos Estados Unidos 154, Bairro das Nac\~oes - Itajub\'a -MG, Brazil.
\\$^3$ Dep. de F\'{i}sica -ICEX - UFMG, Campus da Pampulha, Av. Antonio Carlos 6627, 31270-901 - Belo Horizonte - MG, Brazil.
}
\begin{document}

\date{}
\pagerange{\pageref{firstpage}--\pageref{lastpage}} \pubyear{}

\maketitle

\label{firstpage}

\begin{abstract}
We present SOAR/OSIRIS cross-dispersed near-infrared (NIR) integrated spectra of 12 Galactic globular 
clusters that are  employed to test Maraston (2005, hereafter M05) NIR Evolutionary 
Population Synthesis (EPS) models,
and to provide spectral observational constraints to calibrate future models.
We measured Equivalent Widths (\w) of the most prominent NIR absorption features: $\lambda$ 
1.49\mc, \ion{Mg}{i} $\lambda$ 1.58\mc, \ion{Fe}{i}/\ion{Mg}{i}, $\lambda$ 1.59\mc, 
\ion{Si}{i}, $\lambda$ 1.71\mc, \ion{Mg}{i}, $\lambda$ 2.21\mc, \ion{Na}{i} 
and  $\lambda$ 2.26\mc, \ion{Ca}{i} as well as the $\lambda$ 1.62\mc, $\lambda$  2.29\mc, CO and 
$\lambda$ 2.05\mc, CN molecular bands.  Optical \w\ of  G-band (4300\AA), H$\beta$,
Mg$\rm _2$,  \ion{Fe}{i} (4531\AA, 527\AA\ and 5335\AA), and \ion{Na}{i} (5897\AA) were also measured.
The globular clusters \w\ were compared with model predictions with ages within  4 --- 15 Gyr, and
metallicities between $\frac{1}{200}\,Z\odot$ and 2\,$Z\odot$.
Observed integrated colours ($B-V$, $V-I$ and $V-K_s$) were also compared with models.
The NIR integrated spectra among our sample appear qualitatively similar in most the absorption features. 
The M05 models can properly predict the optical \w\ observed in globular clusters.
Regarding the NIR, they do underestimate the strength of \ion{Mg}{i} 1.49\mc, but they can
reproduce the observed \w\ of  \ion{Fe}{i} 1.58\mc, \ion{Si}{i} 1.59\mc, and CO 2.29\mc,  
in about half of our sample. The remaining objects require the inclusion of intermediate-age 
populations. Thus, we suggest that the presence of C- and O-rich stars in models is important 
to reproduce the observed strengths of metallic lines. Another possibility is the lack of 
$\alpha$-enhancement in the models. In the case of the optical and NIR \ion{Fe}{i} lines, standard models
and those that include blue horizontal branch stars, produce similar results.  
A similar trend is observed for \ion{Na}{i} 5895\AA, while in the case of the G-band, the models with
blue horizontal branch do describe better the observations.  For most of the sample the optical to NIR colours are well
described by the M05 models.In general, M05 models can provide reliable information on the NIR stellar population  of galaxies, but 
only when \w\ and colours are taken together, in other words, \w\ and continuum fluxes should be 
simultaneously fitted.  However, the results should be taken with caution, since the models 
tend to predict results biased towards young ages.

\end{abstract}

%________________________________________________________________
\section{Introduction}

The study of stellar populations is a critical step to understand the  continuum emission of galaxies,  even in
active galactic nuclei (AGN). This occurs because several components such as the non-thermal continuum, dust emission, the presence of an active 
nucleus, and the stellar population itself, sum up to the integrated spectrum. By analysing the stellar 
content, information can be obtained on critical processes like recent star formation episodes, the 
evolutionary history of the galaxy, and the connection between the active galactic nucleus and starburst activity. 

Observations in the near-infrared (NIR) are important to study the integrated stellar populations of galaxies, 
since this is the most convenient spectral region accessible to ground-based telescopes to probe highly obscured sources. 
However, tracking the star formation in the NIR is complicated \citep{origlia00}. Nevertheless, NIR stellar absorption 
features are widely believed to provide means for recognising red supergiants \citep{oliva95}, since they
arise as prime indicators for tracing starbursts in galaxies.
Besides the  short-lived red supergiants, the NIR also includes 
the contribution of thermally- pulsating asymptotic giant
branch (TP-AGB) stars, enhanced in young to intermediate age stellar
populations \citep[$0.2 \leq t \leq 2$ Gyr,][]{maraston98,maraston05,riffel07,riffel08,riffel09}.
The TP-AGB phase becomes fully developed 
in stars with a degenerate carbon oxygen core \citep[see][for a review]{ir83}.
Evidence of this population in the optical is usually missed, as the most
prominent spectral features associated with this  evolutionary phase falls in the NIR \citep[][hereafter M05]{maraston05}.

Usually, the interpretation of a galaxy's  stellar population involves a synthesis approach that, in general, is 
based on the mixing of simple stellar 
populations of different ages and metallicities that provides the best match
to the galaxy spectra. The combinations can be done using equivalent widths (\w ) and selected 
continuum fluxes \citep[e.g.][]{bica86,bica88,riffel08} or by fitting the whole underlying 
spectrum \citep[e.g.][]{cid04,cid05a,cid05b,asari07,cid08,riffel09}. However, in both methods the most 
important ingredient in the SP synthesis is the spectral base. 
An ideal base should cover the  full range of spectral properties that occurs
in a galaxy sample, providing enough resolution in age and metallicity to properly address the desired
scientific question \citep[see][for example]{alex91,cid05a}.
 
A reliable base to probe the stellar population of galaxies is a library of integrated spectra of star clusters 
such as the one constructed by \citet{bica86}.  However this base is restricted to the optical 
region and metallicities lower than solar ($[Z/Z_{\odot}] \lesssim {\rm 0.1}$).
The main advantage of this approach over those based on spectra of individual stars, the so called 
Evolutionary Population Synthesis (EPS) models \citep[][M05]{bc03,vazdekis10}, 
is the reduced number of variables. While the latter are essentially described by 
the temperature, gravity and metallicity, uncertainties in the former are reduced to age and metallicity. 
Also, \citet{bica86} method is free from any assumptions on stellar evolution and the initial 
mass function. However, one advantage of the EPS models is their large range in ages and metallicities, while 
observed spectral libraries are restricted to the optical \citep[e.g.][]{bica86}  or cover only small
fractions of the NIR spectral region, like the region around the 1.6\mc\ and 2.29\mc\ CO bands
\citep{origlia94,origlia97} or only the $K$-band  spectra \citep{lyubenova10}.

The above considerations justify a project to check the reliability of NIR EPS models available 
in the literature and to set important constraints on the absorption features observed in this spectral region. 
We focus here on the observation of NIR ($\sim$1.2\mc - 2.35\mc) integrated spectra of globular clusters.
By their very nature, globular clusters spectra should be reproduced by simple stellar population theoretical models.

This paper is structured as follows: In section \ref{obsdata} observations and data 
reduction procedures are described. Results are presented and discussed in section \ref{results}, and conclusions 
are presented in section \ref{conclusion}.

\section{Sample selection, observations and data reduction}\label{obsdata}

As part of an ongoing project to investigate NIR spectral properties of 
globular clusters, we have selected a representative sub-sample 
of the star clusters from \citet{bica86} to test NIR EPS models. Our original sample
was composed by 27 star clusters covering almost all the possibilities of the age/metallicity that occur in 
galaxies. 

As stated by M05, even bright optical star clusters have 
the NIR light dominated by stars in short-lived evolutionary phases like the bright red 
giant branch (RGB), as well as the TP-AGB.
Thus, to avoid the resolution of individual stars, two strategies were used: (i) the telescope was defocused and 
(ii) non sidereal 
guiding, sweeping the area corresponding to the cluster centre. Consequently, integrated spectra of the mixed light
of a representative fraction of the cluster stars were obtained.

Using these strategies, cross-dispersed (XD) NIR spectra of 23 star clusters were obtained from October 2006 
to June 2009, at the Ohio State Infrared Imager/Spectrometer (OSIRIS) attached to the 4.1\,m Southern 
Astrophysical Research (SOAR) Telescope. The detector is a
1024 $\times$ 1024 HgCdTe array. At the XD mode the spectrograph provides simultaneous coverage 
of the region between 1.2$\mu$m and 2.35$\mu$m with a resolving power of R$\sim$1200. It is interesting 
to note that in this mode it is possible to observe the 1.2$\mu$m to 2.35$\mu$m region in a single shot, avoiding 
the seeing and aperture effects when observing in single bands.

Because of the extended nature of the science targets, it was necessary to take separate sky
exposures, following the pattern ABAB, where A is the on-source observation, with the object centred on
the slit, and B represents off-source exposures. To remove telluric lines and to flux calibrate the spectra, standard stars were 
observed, following the nod pattern along the slit.

Given the nature of the NIR light emitted by globular clusters, the above observational strategy and configuration
produced only 12 reliable (S/N $\gtrsim$ 25) integrated spectra of globular clusters (Tab.\ref{log}). We list 
their observed colours and some basic properties in  Tab.~\ref{prop}. As it can be seen from Tab.~\ref{prop} and  Fig.~1 of \citet{bica06}, seven 
clusters are metal-poor, while the metal-rich ones are almost all located in the Galactic bulge ($\rm d_{GC} \leq 5$kpc).

\begin{table}
\renewcommand{\tabcolsep}{.6mm}
\caption{Observation log.}
\label{log}
\centering
\begin{small}
\begin{tabular}{cccccccc}
\hline
\hline
\noalign{\smallskip}
Object  & Exp. Time & S/N & Date & Object  & Exp. Time &S/N& Date \\
(NGC) &  (s) & & & (NGC)   &  (s) & & \\
\hline
\noalign{\smallskip}
104  & 144 &55 & 10/01/2006 & 6517 & 3120 &65  & 06/10/2009 \\
362  & 170 &45 & 10/01/2006 & 6528 & 1680 &52  & 06/09/2009 \\
1851 & 360 &54 & 10/24/2006 & 6541 & 1920 &43  & 06/10/2009 \\
2808 & 255 &35 & 12/17/2006 & 6553 & 600  &63	& 06/09/2009 \\
6388 & 480 &30 & 06/09/2009 & 6864 & 72   &30	& 10/24/2006\\
6440 & 840 &27 & 06/09/2009 & 7078 & 340  &46	& 10/24/2006\\
\hline
\end{tabular}
\begin{list}{}
\item S/N measured in the $K$-band spectra.
\end{list}
\end{small}
\end{table}

\begin{table*}
%\begin{}
\caption{Observed colours and some basic properties of the cluster sample.}
\label{prop}
\centering
\begin{tabular}{lccccccccccccccc}
\hline
\hline
\noalign{\smallskip}
Data &	l     &  b     & dGC   &  [Fe/H]&   Mv  &   (J-Ks)      & (J-H)	         &  (H-Ks)     & (V-Ks)   & (U-B)  & (B-V) &  (V-R) &  (V-I)  \\
Unit & (deg)  & (deg)  & (kpc) &        & (mag) &  (mag)        &  (mag)         &  (mag)      &  (mag)   & (mag)  & (mag) &  (mag) &  (mag)  \\				
NGC  &  (a)   & (a)    &  (a)  &  (a)   & (b)   &   (c)         &   (c)	         &   (c)       &  (c)     &  (b)   & (b)   &  (b)   & (b)	\\	   
\hline\noalign{\smallskip}
104  & 305.90 & -44.89 &   6.7 &  -0.76 & -9.42 &  0.74\pp0.03  & 0.55\pp0.03	 &  0.19\pp0.04 &   2.60  &  0.37  &  0.88 &   0.53 &	1.14  \\
362  & 301.53 & -46.25 &   8.9 &  -1.16 & -8.40 &  0.64\pp0.03  & 0.52\pp0.03	 &  0.12\pp0.04 &   2.53  &  0.16  &  0.77 &   0.49 &	1.01  \\
1851 & 244.51 & -35.04 &  16.1 &  -1.22 & -8.33 &  0.65\pp0.03  & 0.53\pp0.03	 &  0.12\pp0.04 &   2.66  &  0.17  &  0.76 &   0.49 &	1.01  \\
2808 & 282.19 & -11.25 &  10.5 &  -1.15 & -9.36 &  0.62\pp0.03  & 0.51\pp0.25	 &  0.11\pp0.25 &   2.50  &  0.23  &  0.92 &   0.57 &	1.18  \\
6388 & 345.56 &  -6.74 &   5.0 &  -0.60 & -9.82 &  0.75\pp0.03  & 0.62\pp0.03	 &  0.13\pp0.04 &   2.65  &  0.66  &  1.17 &   0.71 &	1.47  \\
6440 &  7.73  & +3.80  &  1.7  & -0.34  & -8.75 &  0.74\pp0.03  & 0.66\pp0.03	 &  0.08\pp0.04 &   3.06  &  1.47  &  1.97 &   1.24 &	2.51  \\
6517 & 19.23  & +6.76  &  4.7  & -1.37  & -8.28 &  0.49\pp0.03  & 0.48\pp0.03	 &  0.01\pp0.04 &   3.22  &  0.85  &  1.75 &   --   &	2.31  \\
6528 &  1.14  & -4.17  &  2.0  & -0.04  & -6.93 &  0.78\pp0.03  & 0.65\pp0.03	 &  0.13\pp0.04 &   4.04  &  1.09  &  1.53 &   0.9  &	1.74  \\
6541 & 349.29 & -11.18 &   1.9 &  -1.83 & -8.37 &  0.42\pp0.15  & 0.41\pp0.13	 &  0.01\pp0.19 &   2.33  &  0.13  &  0.76 &   0.49 &	1.01  \\
6553 &  5.25  & -3.02  &  1.7  & -0.21  & -7.99 &  0.87\pp0.03  & 0.69\pp0.03	 &  0.18\pp0.04 &   3.73  &  1.34  &  1.73 &   1.01 &	2.13  \\
6864 & 20.30  &-25.75  & 14.6  & -1.16  & -8.35 &  0.61\pp0.03  & 0.52\pp0.03	 &  0.09\pp0.04 &   2.36  &  0.28  &  0.87 &   0.55 &	1.16  \\
7078 & 65.01  &-27.31  & 10.1  & -1.62  & -9.17 &  0.58\pp0.00  & 0.47\pp0.00	 &  0.11\pp0.00 &   2.16  &  0.06  &  0.68 &   --   &	0.85  \\
\hline
\end{tabular}
\begin{list}{Table Notes:}
\item (a) From \citet{bica06}; (b) From - http://www.physics.mcmaster.ca/Globular.html \citep{harris96}; 
(c) From \citet{cohen07}. 
\item The NIR colours are reddening-corrected \citep{cohen07}
while the optical ones are uncorrected \citep{harris96}.
\end{list}
%\end{tiny}
\end{table*}

% Data Reduction
The spectral reduction, extraction and wavelength calibration procedures were performed 
using XD-Spres\footnote{http://www.if.ufrgs.br/$\sim$riffel/software.html.} {\sc iraf} software task (Ruschel-Dutra et al. 2010, {\it in preparation}).
This tool follows the standard NIR cross dispersed reduction procedures, as listed:
(i) to remove sky emission lines from the integrated spectra, sky exposures were subtracted from the object. 
(ii) the resulting images were summed up and then divided by a normalised flat-field image.
(iii) extraction was performed following the standard echelle procedures.
(iv) wavelength calibration was based on the  OH lines present in the sky exposures with the values given by \citet{oliva92}. 
(v) the 1-D spectra were then corrected for telluric absorption by comparison with an A0V star spectrum. The stellar atmospheric absorption lines 
were identified and removed. This procedure was done using the {\sc telluric} task. 
(vi) finally, the flux calibration was achieved by fitting a black-body curve to the standard 
star featureless spectrum, using the task {\sc calibrate} of the {\sc iraf} software.  
Final reduced spectra, are shown in Fig.~\ref{fullspec}.

\section{Results and Discussion}\label{results}

\subsection{General overview of the spectra}

\begin{figure*}
\centering
\includegraphics[]{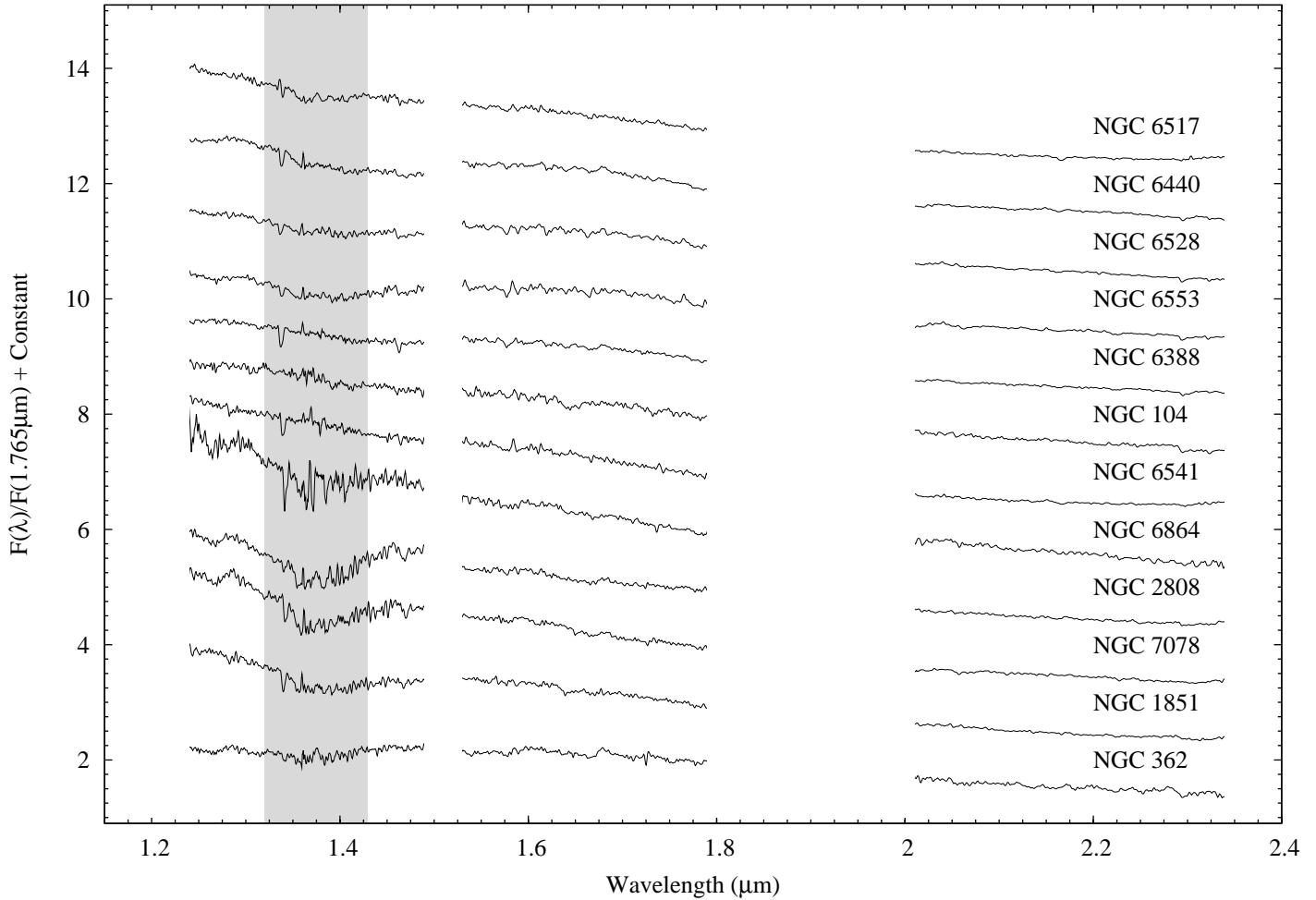}
\caption{Globular Clusters NIR spectral energy distribution. The spectra were normalised by the mean value of their fluxes between 
1.76\mc\ and 1.77\mc and shifted by a constant for display purposes. The shaded region was left out of the analysis (see text). 
The gaps represent regions strongly dominated by telluric absorptions, and/or coincide with band limits. }
\label{fullspec}
\end{figure*}

Spectra of the 12 Galactic globular clusters covering simultaneously the NIR 
spectral region between 1.2\mc\ - 2.35\mc\ are presented here for the first time. The 
full NIR spectral energy distribution (SED) of the clusters is shown in Fig.~\ref{fullspec}. 
The NIR SED of the clusters is rather similar among the 12 objects analysed, decreasing 
smoothly in flux with wavelength, and present similar absorption features. The shaded area in 
Fig.~\ref{fullspec} denotes a region highly affected by telluric absorption, thus it was left out
of further analysis.

\begin{figure*}
\begin{minipage}[b]{0.45\linewidth}
\centering
\includegraphics[scale=0.7]{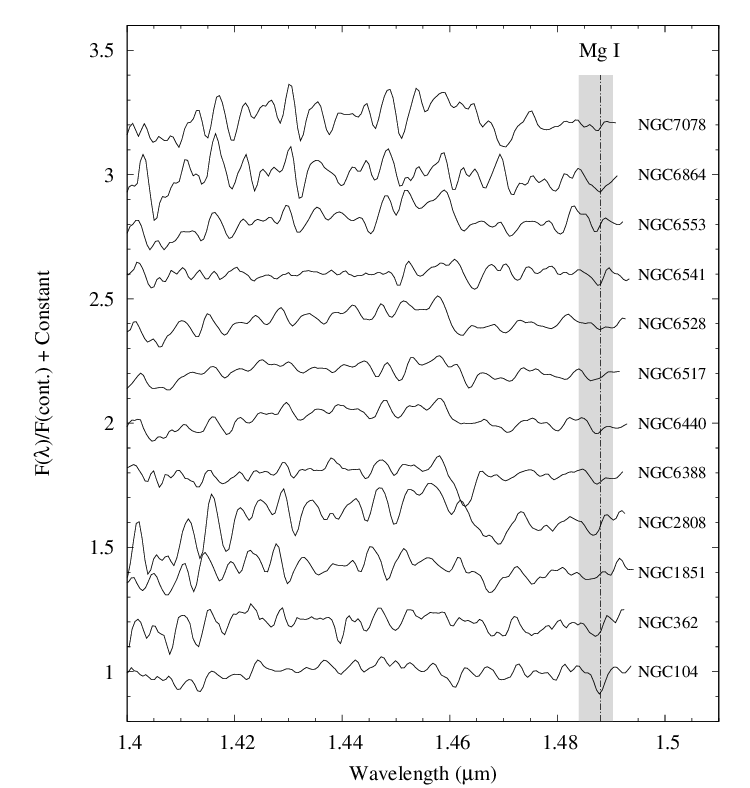}
\caption{J-band final reduced spectra. The most prominent lines are identified. The shaded region represents 
the line limits and the vertical line shows the line centre.}
\label{spectra-j}
\end{minipage}
\hspace{0.5cm}
\begin{minipage}[b]{0.45\linewidth}
\centering
\includegraphics[scale=0.7]{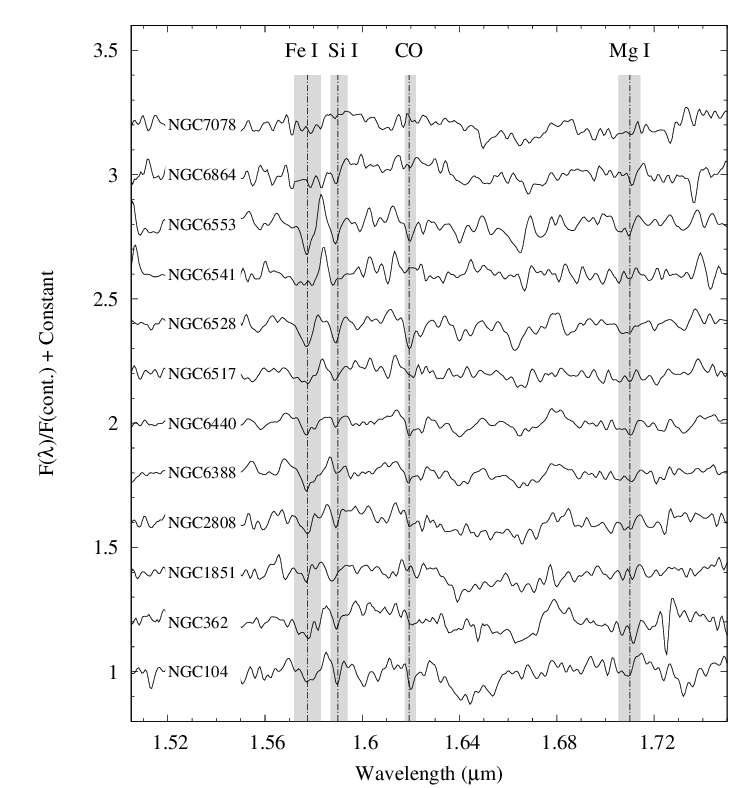}
\caption{H-band final reduced spectra. The most prominent lines are identified. The shaded region represents 
the line limits and the vertical line shows the line centre.}
\label{spectra-h}
\end{minipage}
\end{figure*}

\begin{figure*}
\begin{minipage}[b]{0.45\linewidth}
\centering
\includegraphics[scale=0.7]{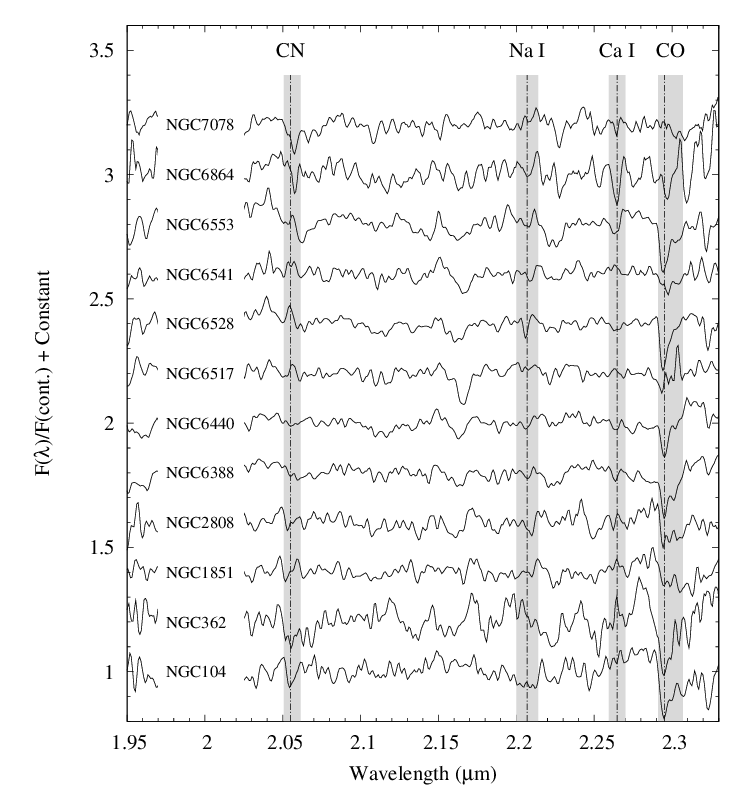}
\caption{K-band final reduced spectra. The most prominent lines are identified. The shaded region represents 
the line limits and the vertical line shows the line centre.}
\label{spectra-k}
\end{minipage}
\hspace{0.5cm}
\begin{minipage}[b]{0.45\linewidth}
\centering
\includegraphics[scale=0.45]{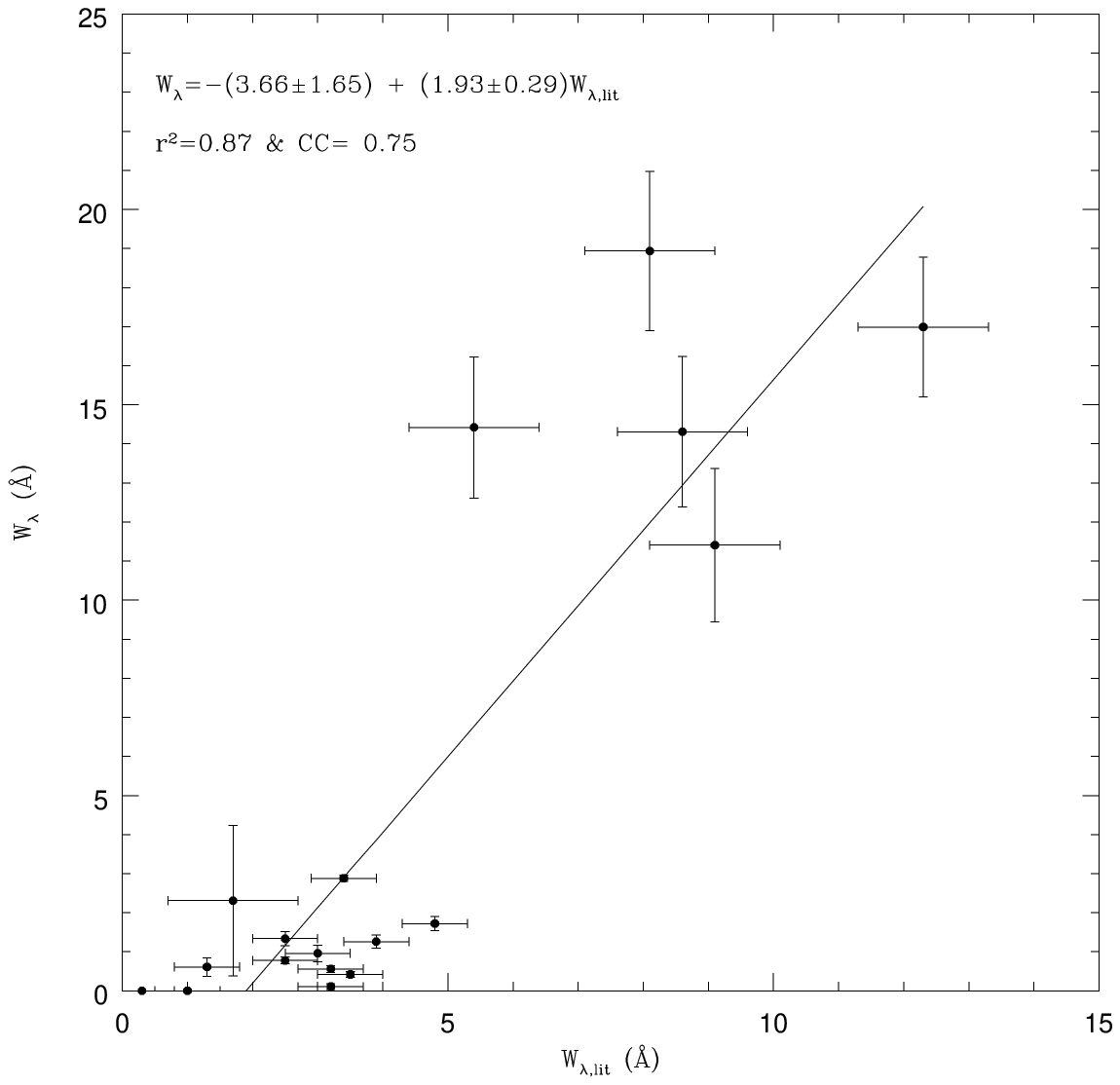}
\caption{Comparison of the measured \w\ with literature data from \citet{origlia97}. Features used are 
1.59\mc\ (Si {\sc i}), 1.62\mc\ (CO 6-3), 2.29\mc\ (CO 2-0). The Pearson coefficient ($\rm r^2$),
a generalised correlation coefficient (CC, see text) as well as the regression equation are shown.}
\label{lit}
\end{minipage}
\end{figure*}

In order to highlight the absorption lines, the spectra were linearised by their continuum. The 
result of this process is shown in Figs.~\ref{spectra-j} to \ref{spectra-k}. Clearly,
the spectra of the globular cluster sample appear qualitatively similar in most of the NIR absorption features. 
Such similarity can be observed in Figs.~\ref{spectra-j} to \ref{spectra-k}, where many atomic absorption 
features like $\lambda$ 1.49\mc\ \ion{Mg}{i}, $\lambda$ 1.58\mc\ Fe {\sc i}/\ion{Mg}{i}, $\lambda$ 1.59\mc\ 
Si {\sc i}, $\lambda$ 1.71\mc\ Mg {\sc i}, $\lambda$ 2.21\mc\ Na {\sc i}
and  $\lambda$ 2.26\mc\ \ion{Ca}{i} as well as the $\lambda$ 1.62\mc, $\lambda$  2.29\mc\ CO and 
$\lambda$ 2.05\mc\ CN molecular bands are clearly detected and identified in the spectra.  The centre 
and width of the strongest lines are identified in Figs.~\ref{spectra-j} to \ref{spectra-k}. The similarity 
between the spectra is not surprising, since the sample is composed only by globular clusters and thus, the 
integrated spectra is dominated by stars of relatively similar spectral type.

It is also clear in Figs.~\ref{spectra-j} to \ref{spectra-k}  that the spectra show many weak absorption 
lines. However, as the S/N (see Tab.~\ref{log}) is relatively poor, no effort was made to identify such weak absorptions.
In addition, the $H$ and $K$-band lines presently detected are common in the brightest stars of star 
clusters \citep{origlia02,origlia06,frogel01,stephens04}, which indicates that our data reduction is consistent.

To date the only effort in obtaining NIR integrated spectra of Galactic star clusters was from \citet{origlia94,origlia97}.  
These authors used the IRSPEC infrared spectrometer \citep{moorwood91} attached to the European
Southern Observatory (ESO) New Technology Telescope (NTT). Their instrumental setup allowed them to obtain 
long-slit spectra centred at 1.59\mc\ (Si {\sc i}), 1.62\mc\ (CO 6-3), and 2.29\mc\ (CO 2-0). These
absorptions are clearly detected in our spectra (Figs.~\ref{spectra-j} - \ref{spectra-k}). 
 For a comparison between both data sets, the \w\ reported by \citet{origlia94,origlia97} for the objects 
in common with our sample were plotted 
against those measured by us (Fig.~\ref{lit}). 
The Pearson correlation coefficient ($\rm r^2=0.87)$ and a nonlinear regression were computed
using the {\sc fitexy} routine from the IDL Astronomy User's Library, which takes errors in both coordinates into account.
We find $\rm W_{\lambda}=-(3.66\pm1.65)+(1.93\pm0.29)\,W_{\lambda,lit}$, where \w\ means this work and 
$W_{\lambda,lit}$ the literature data.  A generalised correlation coefficient (CC=0.75), defined as the square 
root of the fraction of total $y$ variance explainable by the fitted function was also computed.
Clearly, there is a discrepancy between our 
values and those of \citet[][]{origlia97}. Probably, the difference in \w\ of Si {\sc i} 1.59\mc\ and CO 1.62\mc\ 
occurs because these authors use rectified spectra, instead of locally defined continuum regions. For the 
2.29\mc\ CO band, the differences in \w\ are associated with different 
line definitions (see Sec.~\ref{equvalentewidths}). In addition, the difference may be partly
due to the fact that \citet{origlia94,origlia97} only sampled a very small fraction of the cluster core ($6\arcsec\times4\arcsec$).  Thus,
the random contribution of a few bright, red giant stars in such a small field, may strongly affect features in the integrated spectra.

\subsection{Equivalent Widths \label{equvalentewidths}}

A straightforward approach for comparing a set of empirical spectra with models is by means of the \w\ of their
absorption features. Such a procedure is virtually free from flux calibration issues and reddening corrections, and 
may provide constraints on model predictions.

\begin{figure}
\includegraphics[scale=0.55]{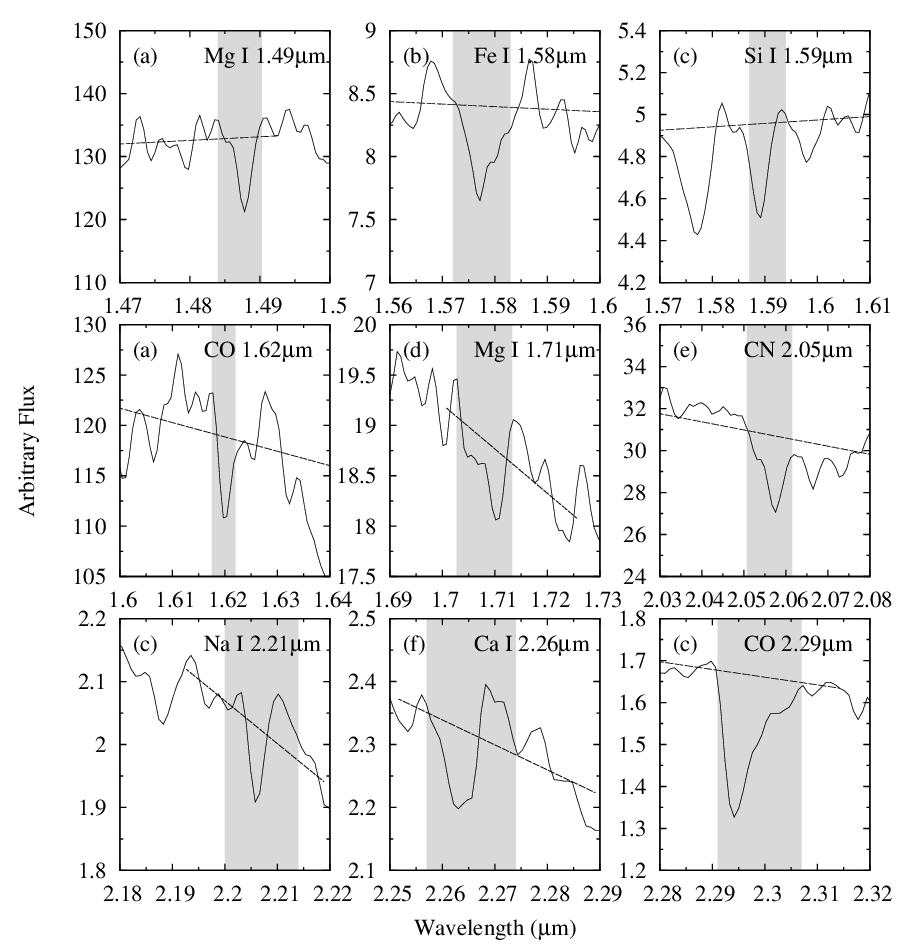}
\caption{Examples of line limits and continuum band-passes. The labels (a), (b), (c), (d), (e)
and (f) indicate the cluster from which they were taken, being: NGC 104, NGC 6388, NGC 6528, NGC 6440, NGC 7078 and NGC 6553, respectively.}
\label{bandpass}
\end{figure}

When dealing with optical absorption lines, the spectral window of a feature is 
usually defined together with two continuum regions at the red and blue sides, which are  used to trace 
a local continuum through  a linear fit to the mean values of both continuum regions  \citep[e.g.][]{faber85,bica86}.
Similar definitions were made for some NIR absorption features \citep[e.g.][]{oliva95,frogel01}, however, 
they were based in stellar spectra, where absorption features are typically a few \kms\ wide.  As we are dealing  with 
velocity-dispersion sustained ($\sigma_v \gtrsim$10\kms) clusters, and we aim to use these absorption features to investigate 
unresolved stellar populations in galaxies, we are forced to use spectral indices with broader band passes than those used in individual stars. 
Our continuum and band-pass definitions are listed in Tab.~\ref{ewdefs}.  They were taken from the literature, except for
\ion{Mg}{i} 1.49\mc\ and CN 2.05\mc, which we define in Tab.~\ref{ewdefs}. In some cases it was also necessary to re-define the continuum regions in 
to avoid overlap (Tab.~\ref{ewdefs}).  However, special care was taken to use only regions 
free from emission/absorption lines. An example of these continuum and line limits is shown in Fig.~\ref{bandpass}.

\begin{table}
\renewcommand{\tabcolsep}{1.5mm}
\caption{Line limits and continuum band-passes.}
\label{ewdefs}
%\begin{tiny}
\begin{tabular}{lcccc}
\hline
\hline
\noalign{\smallskip}
Line                & $\rm Ew_{\lambda b}$ & $\rm Ew_{\lambda r}$ & $\rm CBP_{b}$ & $\rm CBP_{r}$  \\
\hline
\noalign{\smallskip}
\ion{Mg}{i} 1.49\mc\     & 1.4840   & 1.4903     & 1.4620 - 1.4830	   & 1.4910 -  1.4925  	    \\
\ion{Fe}{i} 1.58\mc\ (a) & 1.5720   & 1.5830     & 1.5400 - 1.5700	   & 1.5950 - 1.6160	    \\
\ion{Si}{i} 1.59\mc\ (b) & 1.5870   & 1.5910     & 1.5400 - 1.5700	   & 1.5950 - 1.6160	    \\
CO          1.62\mc\ (b) & 1.6175   & 1.6220     & 1.5950 - 1.6160	   & 1.6280 - 1.6570	    \\
\ion{Mg}{i} 1.71\mc\ (c) & 1.7053   & 1.7143     & 1.7010 - 1.7056	   & 1.7156 - 1.7256	    \\
CN          2.05\mc\     & 2.0507   & 2.0615     & 2.0250 - 2.0350         & 2.0700 - 2.1730	    \\
\ion{Na}{i} 2.21\mc\ (c) & 2.2000   & 2.2140     & 2.1934 - 2.1996	   & 2.2150 - 2.2190	    \\
\ion{Ca}{i} 2.26\mc\ (c) & 2.2594   & 2.2700     & 2.2516 - 2.2590	   & 2.2716 - 2.2888	    \\
CO          2.29\mc\ (c) & 2.2910   & 2.3070     & 2.2716 - 2.2888	   & 2.3120 - 2.3140        \\
\hline
\end{tabular}
\begin{list}{Table Notes:}
\item CBP$\rm _b$ and CBP$\rm _r$ are the blue and red bandpass boundaries for the continuum. 
(a) from \citet{riffel08}, (b) from \citet{origlia93}, (c) from \citet{cesetti09}.
\end{list}
%\end{tiny}
\end{table}

With the definitions for continuum and band pass listed in Tab.~\ref{ewdefs}, \w\ was measured for nine 
absorption lines, using the code {\sc pacce}, which computes \w\ with user definitions for the 
line and continuum band-passes \citep{vale07}. 
The measured \w\  values are 
presented in Tab.~\ref{ewmeas}.  Errors were estimated according to Eq.~7 of \citet{vollman06}.

In addition, we measured on the 12 globular clusters, \w\ using the Lick indices 
definitions \citep[e.g.][taken from: http://astro.wsu.edu/worthey/html/system.html]{whortey94} for the following optical 
features:  $G$-band (4300\AA), \ion{Fe}{i} (4531\AA, 5270\AA\ and 5335\AA), H$\beta$, MgH (5176\AA), and \ion{Na}{i} (5895\AA),  
MgH (5176\AA). The \w\ corresponding to the MgH absorption is called Mg$\rm _2$, which is adopted in the
present paper. The optical integrated spectra of the clusters were taken from \citet{santos02}.  Observation and reduction 
procedures, as well as data quality of the optical spectra, are discussed in \citet{bica86}. The optical 
\w\ are listed in Tab.~\ref{ewmeas}. Errors were estimated similarly to the NIR \w.

\begin{table*}
\renewcommand{\tabcolsep}{0.65mm}
\begin{scriptsize}
\caption{Measured \w\ (in \AA) for the cluster sample.}
\label{ewmeas}
\centering
\begin{tabular}{lccccccccccccccccccccccc}
\hline
\hline
\noalign{\smallskip}
\w\ & G &           \ion{Fe}{i}   & H$\beta$    & Mg$\rm _2$   & \ion{Fe}{i} &  \ion{Fe}{i}&\ion{Na}{i} &  Mg {\sc i}  &   Fe {\sc i}  & Si {\sc i}    &   CO          &  Mg {\sc i}    &  CN          & Na {\sc i}    &  Ca {\sc i}    &   CO         \\
NGC  & 4300\AA      & 4531\AA     &  4863        & 5176\AA      &    5270     &   5335      &5895\AA    &1.49\mc\   &  1.58\mc\     &  1.59\mc\     &   1.62\mc\    &   1.71\mc\     &   2.05\mc\   &  2.21\mc\     &   2.26\mc\     &  2.29\mc\  \\
\noalign{\smallskip}
\hline
\noalign{\smallskip}
104  &   3.50\pp0.24 &   1.53\pp0.36 &   1.33\pp0.18 &   5.26\pp0.04 &   1.09\pp0.16 &   1.13\pp0.50 &   1.73\pp0.16 & 1.59\pp0.31 & 1.41\pp0.25 &   0.56\pp0.09 &   1.34\pp0.18 &   0.58\pp1.05 &     ---     &   7.05\pp3.48 &       ---     &   18.94\pp2.04 \\
362  &   2.68\pp0.23 &   1.34\pp0.36 &   1.56\pp0.18 &   1.33\pp0.04 &   0.72\pp0.16 &   0.83\pp0.51 &   0.63\pp0.16 & 1.05\pp0.32 & 5.98\pp0.24 &   1.38\pp0.09 &   0.74\pp0.22 &   1.97\pp1.22 & 7.10\pp0.16 &       ---     &       ---     &   23.32\pp1.89 \\
1851 &   1.50\pp0.23 &   0.85\pp0.37 &   1.57\pp0.18 &   1.38\pp0.04 &   1.09\pp0.16 &   0.52\pp0.51 &   0.62\pp0.16 & 1.73\pp0.31 & 0.64\pp0.24 &   1.08\pp0.08 &	 ---	 &	 ---	 &     ---     &   0.52\pp4.61 &       ---     &   4.46\pp2.07 \\ 
2808 &   2.46\pp0.23 &   2.00\pp0.36 &   1.83\pp0.18 &   0.96\pp0.04 &   0.98\pp0.16 &   0.60\pp0.51 &   1.47\pp0.16 & 1.15\pp0.34 & 2.30\pp0.24 &   0.78\pp0.09 &   0.61\pp0.24 &	 ---	 & 0.25\pp0.15 &   4.18\pp4.19 &   0.10\pp0.93 &   14.42\pp1.81 \\
6388 &   2.70\pp0.23 &   2.14\pp0.36 &   1.31\pp0.18 &   4.91\pp0.04 &   1.23\pp0.17 &   1.18\pp0.50 &   3.43\pp0.15 & 1.79\pp0.31 & 4.69\pp0.23 &   $>$0.11     &   0.96\pp0.21 &   0.62\pp1.05 & 2.93\pp0.15 &       ---     &   0.60\pp0.96 &   14.31\pp1.93 \\
6440 &   3.11\pp0.25 &   2.24\pp0.36 &   1.81\pp0.17 &   6.86\pp0.07 &   0.90\pp0.17 &   0.91\pp0.51 &   3.68\pp0.15 & 0.13\pp0.32 & 2.34\pp0.22 &   0.42\pp0.08 &   1.26\pp0.17 &   1.07\pp1.05 & 0.58\pp0.12 &   1.59\pp3.83 &   0.52\pp1.00 &   11.41\pp1.96 \\
6517 &   2.62\pp0.23 &   3.39\pp0.38 &   1.79\pp0.18 &       ---     &   0.97\pp0.16 &   0.85\pp0.51 &   2.42\pp0.15 & 0.92\pp0.32 & 3.22\pp0.22 &   1.36\pp0.08 &   0.69\pp0.18 &	 ---	 & 0.59\pp0.13 &       ---     &       ---     &   2.12\pp1.87 \\ 
6528 &   3.04\pp0.28 &   3.04\pp0.36 &   0.58\pp0.18 &   8.07\pp0.05 &   1.71\pp0.17 &   1.72\pp0.50 &   4.39\pp0.15 & 1.36\pp0.30 & 5.14\pp0.23 &   2.67\pp0.08 &   2.89\pp0.21 &   1.93\pp1.03 &     ---     &       ---     &   2.32\pp0.87 &   14.78\pp1.76 \\
6541 &   1.96\pp0.23 &       ---     &   2.09\pp0.17 &   0.94\pp0.04 &   0.99\pp0.16 &   0.89\pp0.50 &   1.46\pp0.16 & 0.80\pp0.31 & 4.07\pp0.24 &   1.74\pp0.09 &	 ---	 &	 ---	 &     ---     &   0.64\pp3.62 &       ---     &   9.85\pp1.86 \\ 
6553 &   9.95\pp0.33 &   1.94\pp0.45 &   1.71\pp0.19 &   8.27\pp0.06 &   2.71\pp0.17 &   1.73\pp0.50 &   3.50\pp0.16 &     ---     & 5.93\pp0.22 &   2.88\pp0.08 &   1.72\pp0.18 &   1.70\pp1.04 & 4.85\pp0.13 &       ---     &   2.09\pp0.90 &   16.99\pp1.79 \\
6864 &   1.26\pp0.24 &   0.73\pp0.37 &   2.49\pp0.17 &   2.56\pp0.04 &   0.87\pp0.16 &   0.50\pp0.51 &   0.78\pp0.16 & 0.72\pp0.30 & 5.00\pp0.22 &   1.71\pp0.08 &	 ---	 &   0.78\pp1.06 & 0.64\pp0.19 &       ---     &   2.64\pp0.90 &   3.76\pp1.95 \\ 
7078 &       ---     &   0.21\pp0.37 &   1.94\pp0.18 &   0.57\pp0.04 &   0.10\pp0.17 &   0.10\pp0.51 &   1.49\pp0.16 &     ---     & 1.60\pp0.22 &	 ---	 &	 ---	 &	 ---	 & 6.24\pp0.13 &       ---     &       ---     &   2.31\pp1.93 \\ 
\hline
\end{tabular}
\end{scriptsize}
\end{table*}

\subsection{Observations compared to models}

Theoretical spectral  models for SSPs have become fundamental in studies 
of the stellar population of galaxies. Thus, as more such models are available, it becomes 
increasingly important to test them over all the spectral wavelengths \citep{cid10}. 
Our main goal in this paper is to compare the integrated spectra of actual globular clusters with theoretical predictions of 
M05 models. These models were chosen due to the fact that up to date they are the only ones
that make predictions about the presence of almost all observed absorption lines and bands in the NIR spectral region.

%\subsubsection{Absorption features}

In order to compare observations with models we measured in M05 
models the same set of \w\ (using our continuum and line definitions) as in the star clusters. The \w\ of the 
optical absorption features were also measured on the models using the definitions described 
in Sec.~\ref{equvalentewidths}.

%%%%%%%%%%%%%%%%%%%%%%%%%% Equivalent Width %%%%%%%%%%%%%%%%%%%%%%%%%%

\begin{figure*}
\begin{minipage}[b]{0.33\linewidth}
\includegraphics[scale=0.30]{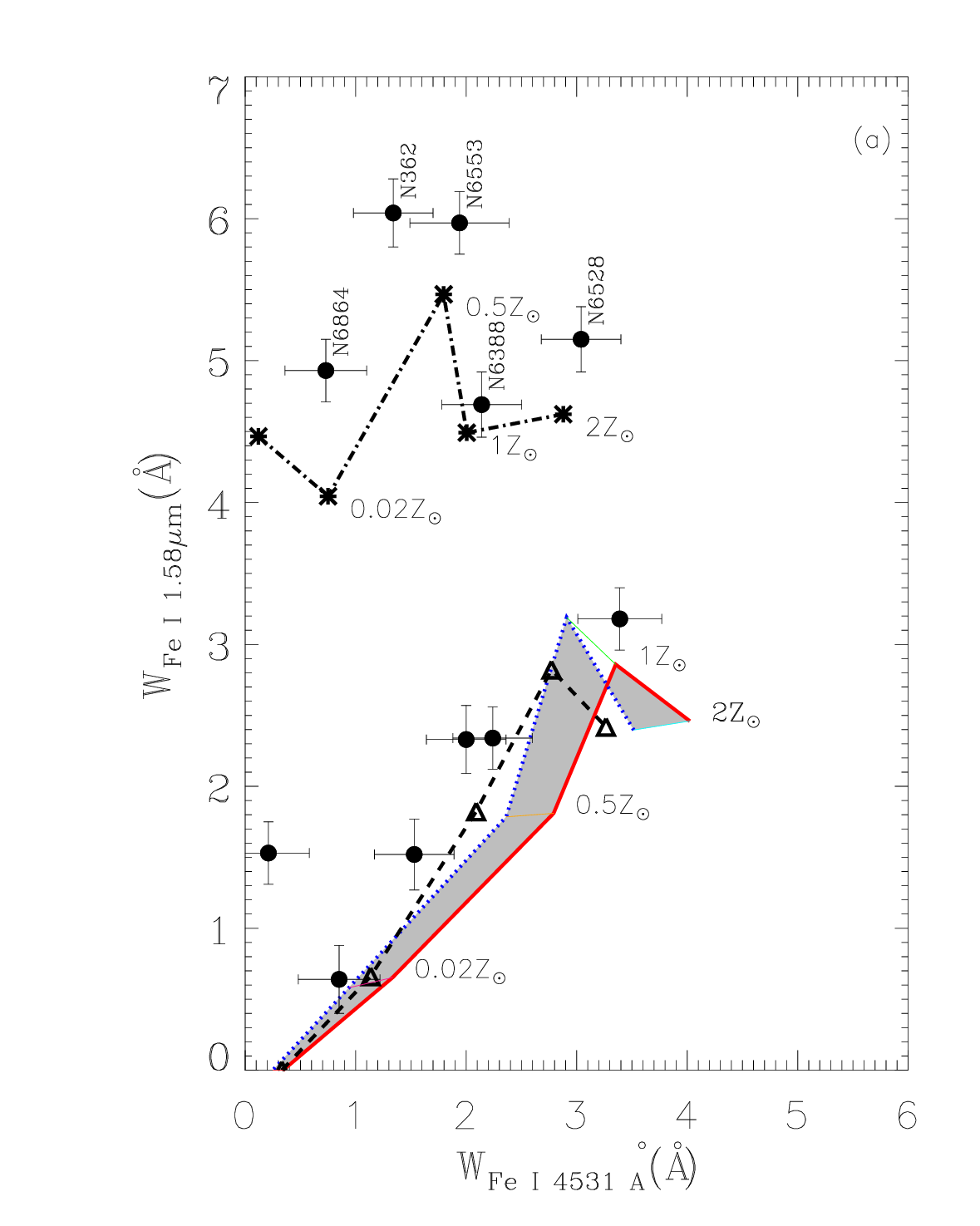} %a
\end{minipage}
\hfill
\begin{minipage}[b]{0.33\linewidth}
\includegraphics[scale=0.30]{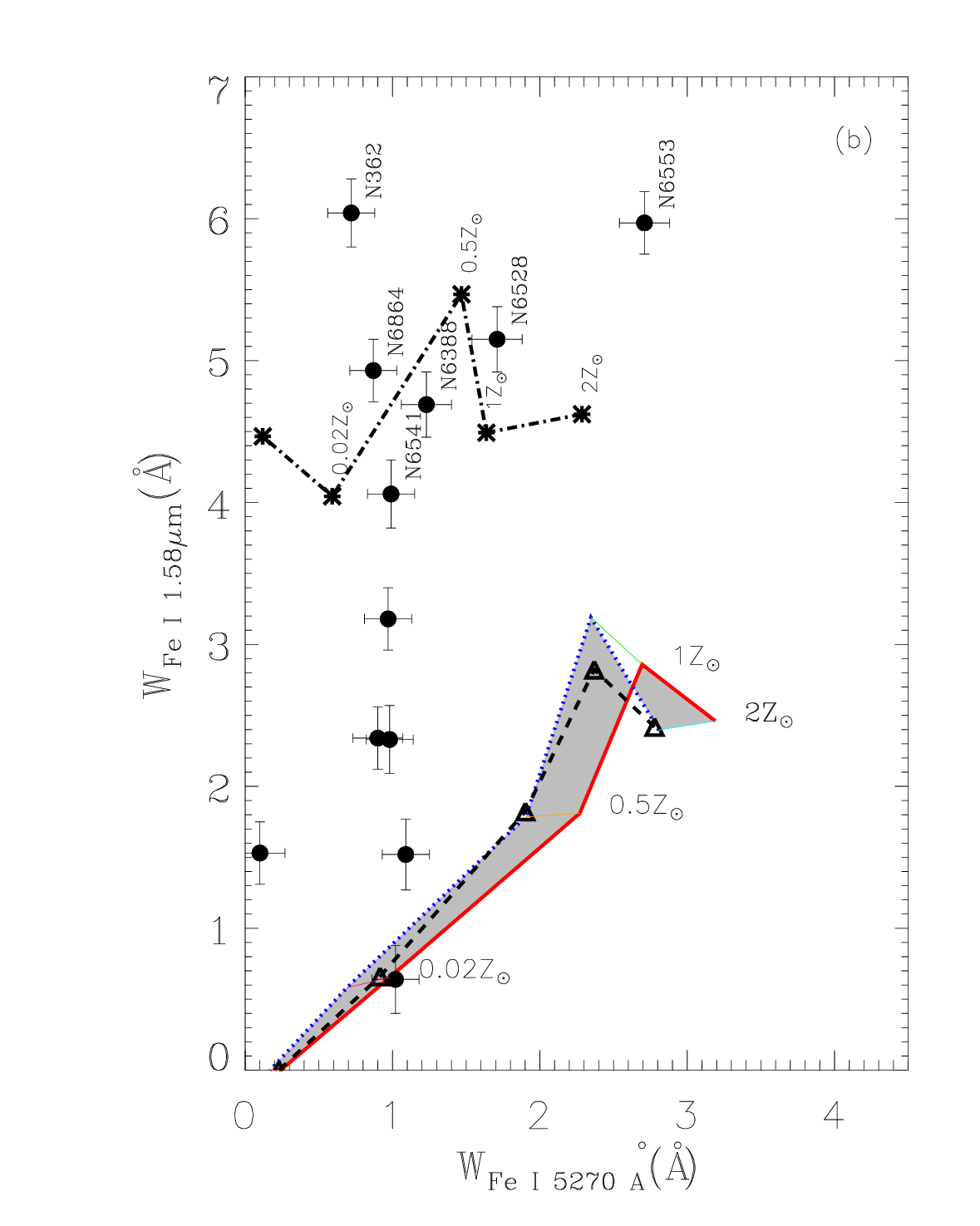} %b
\end{minipage}
\begin{minipage}[b]{0.33\linewidth}
\includegraphics[scale=0.30]{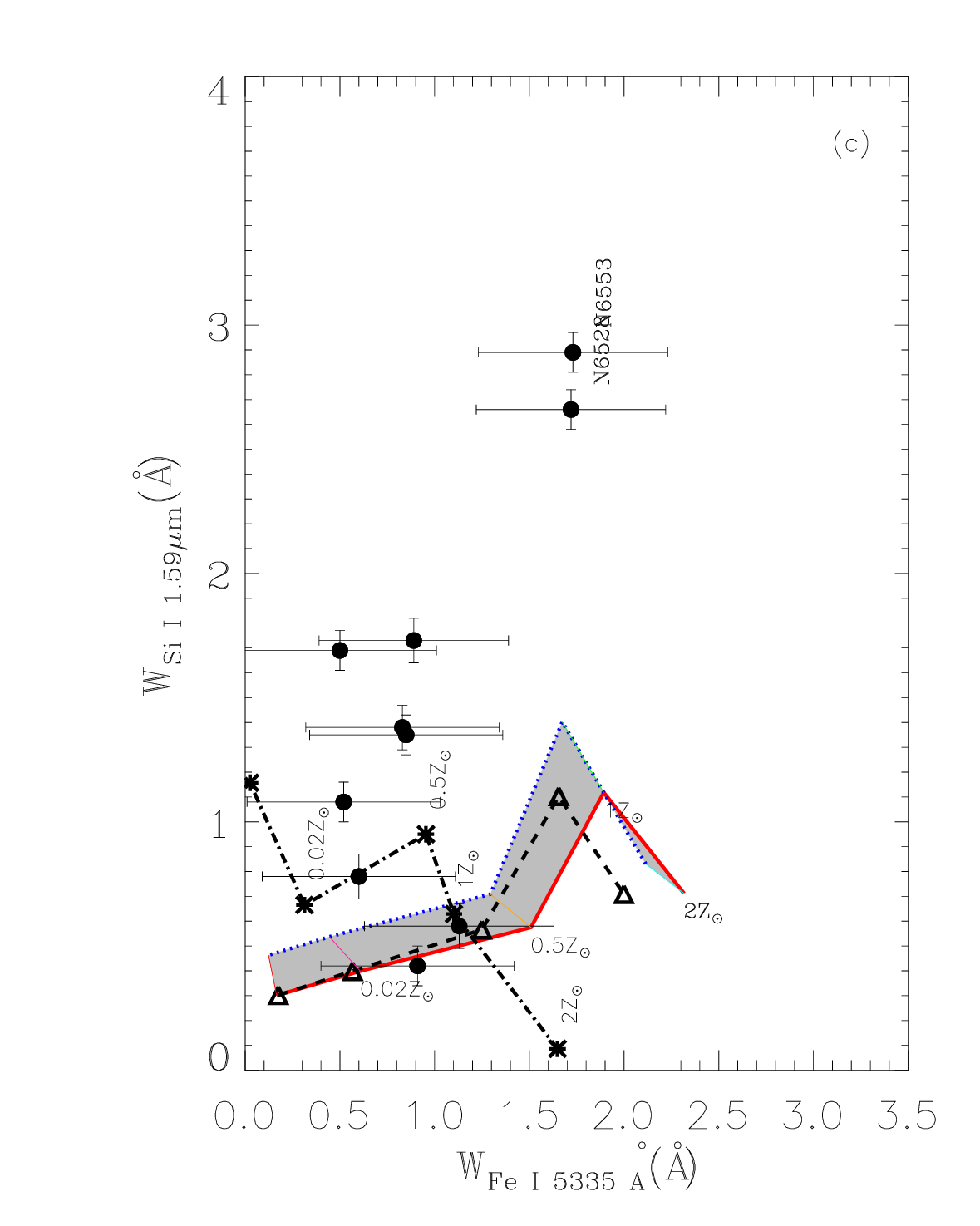} %c
\end{minipage}
\hfill
\begin{minipage}[b]{0.33\linewidth}
\includegraphics[scale=0.30]{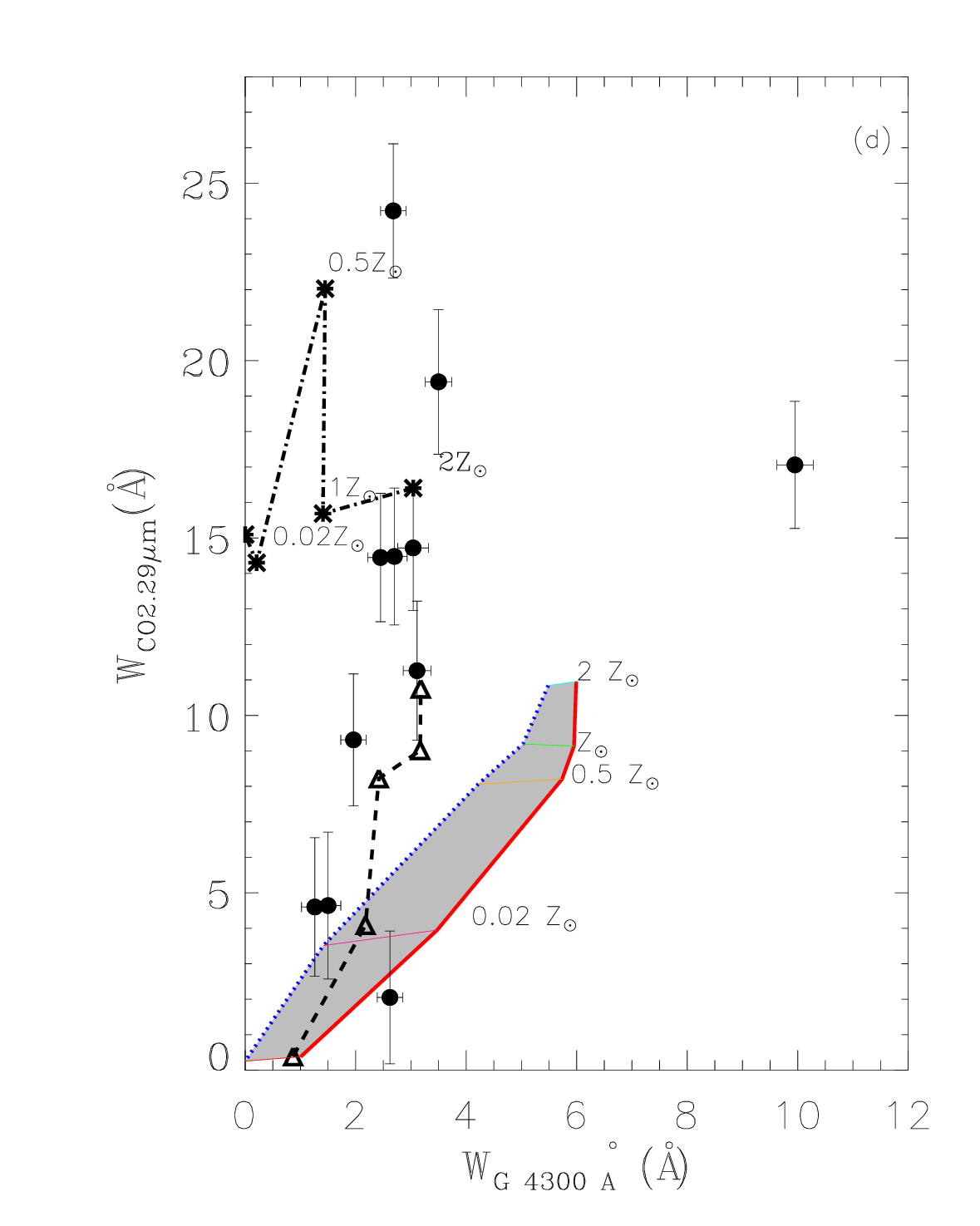} %d
\end{minipage}
\begin{minipage}[b]{0.33\linewidth}
\includegraphics[scale=0.30]{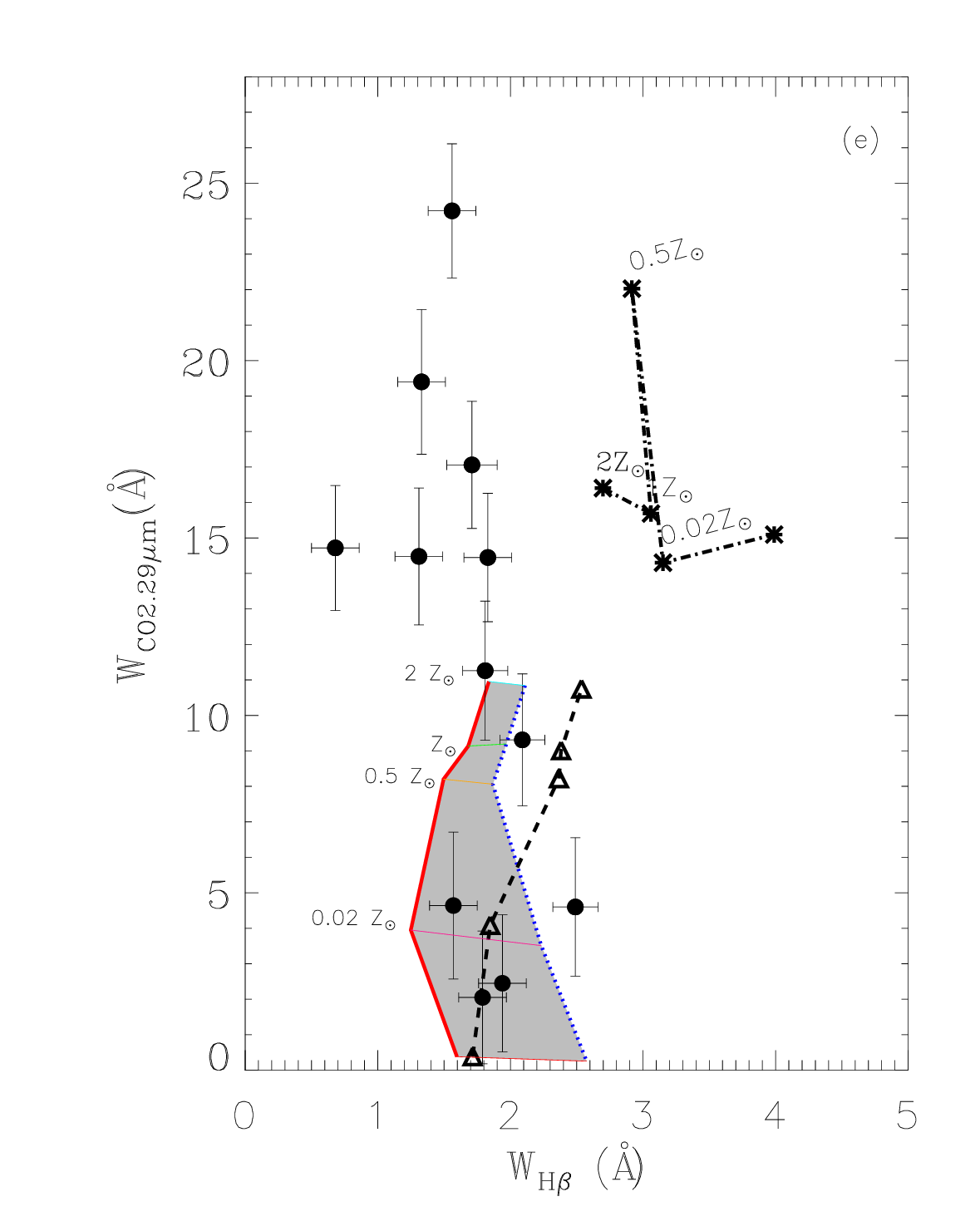} %e
\end{minipage}
\hfill
\begin{minipage}[b]{0.33\linewidth}
\includegraphics[scale=0.30]{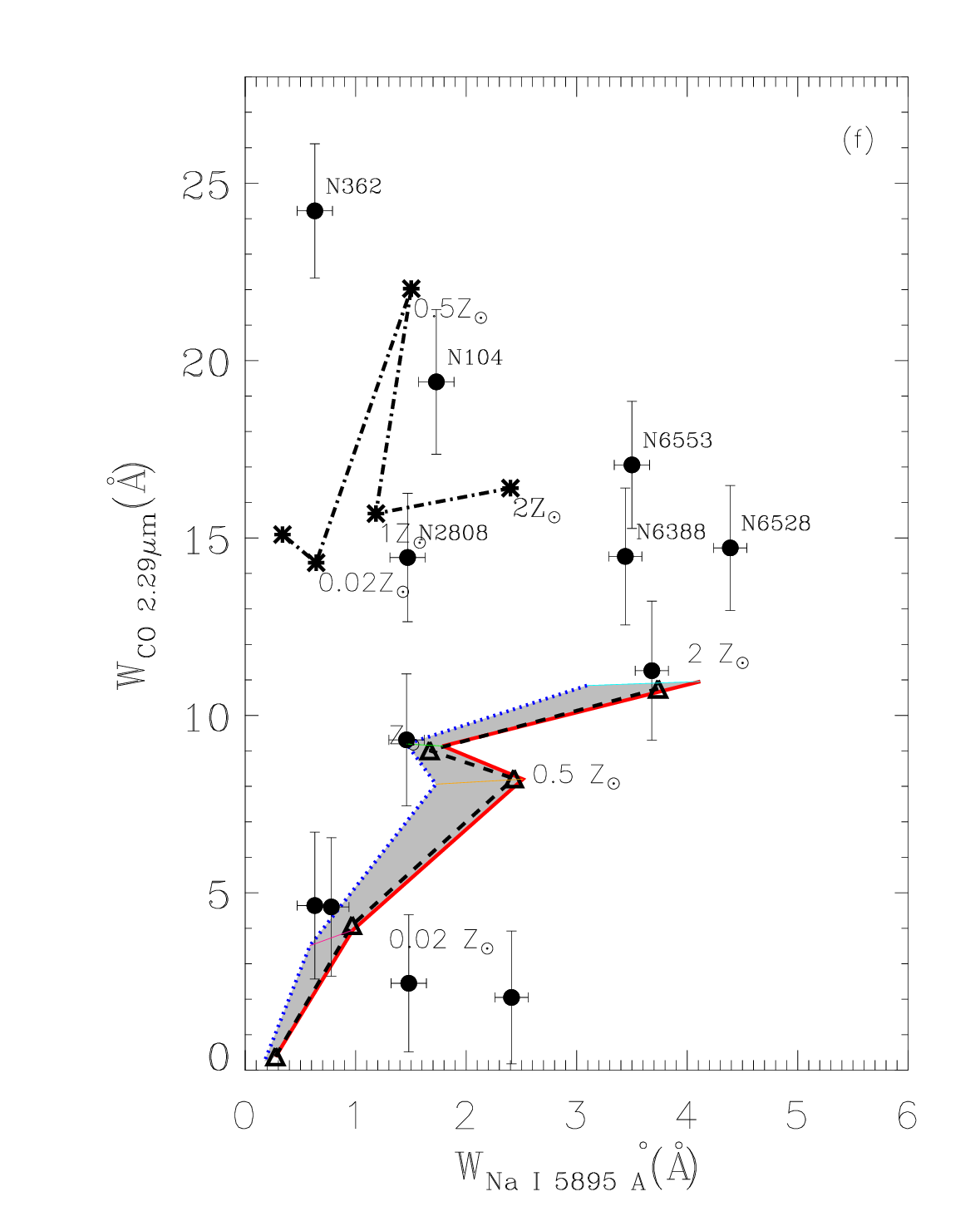} %f
\end{minipage}
\hfill
\begin{minipage}[b]{0.33\linewidth}
\includegraphics[scale=0.30]{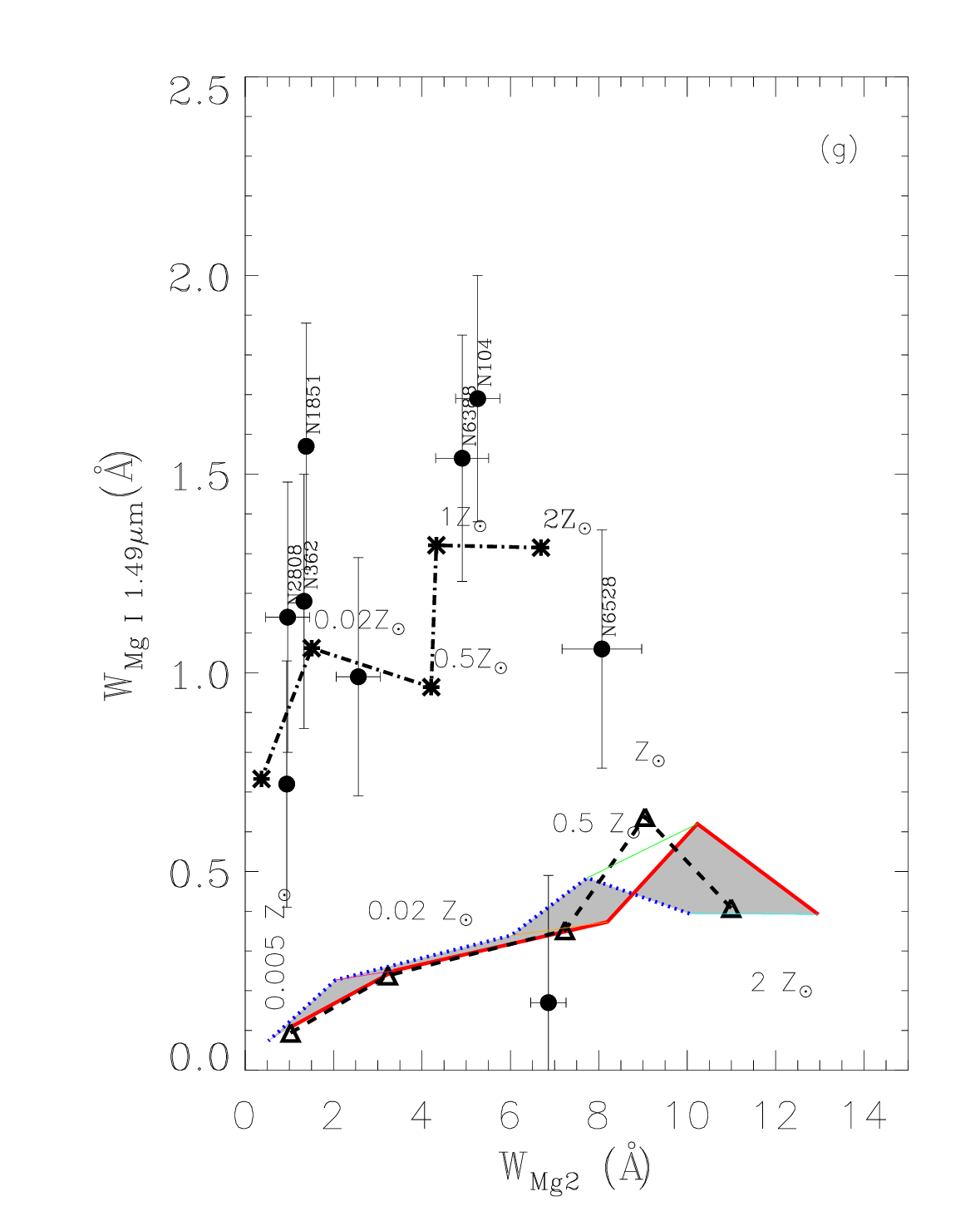} %g
\end{minipage}
\begin{minipage}[b]{0.33\linewidth}
\includegraphics[scale=0.30]{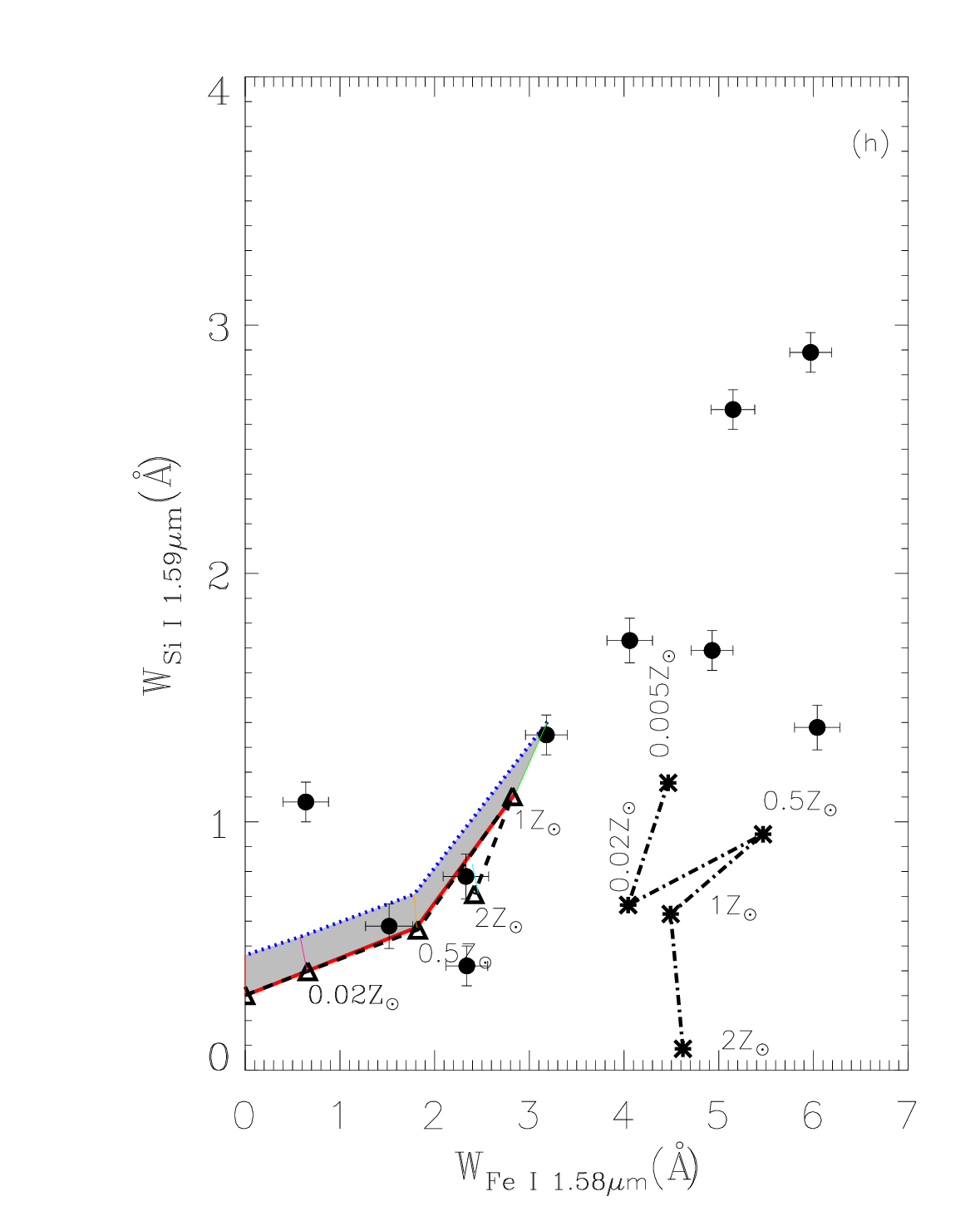} %h
\end{minipage}
\hfill
\begin{minipage}[b]{0.33\linewidth}
\includegraphics[scale=0.30]{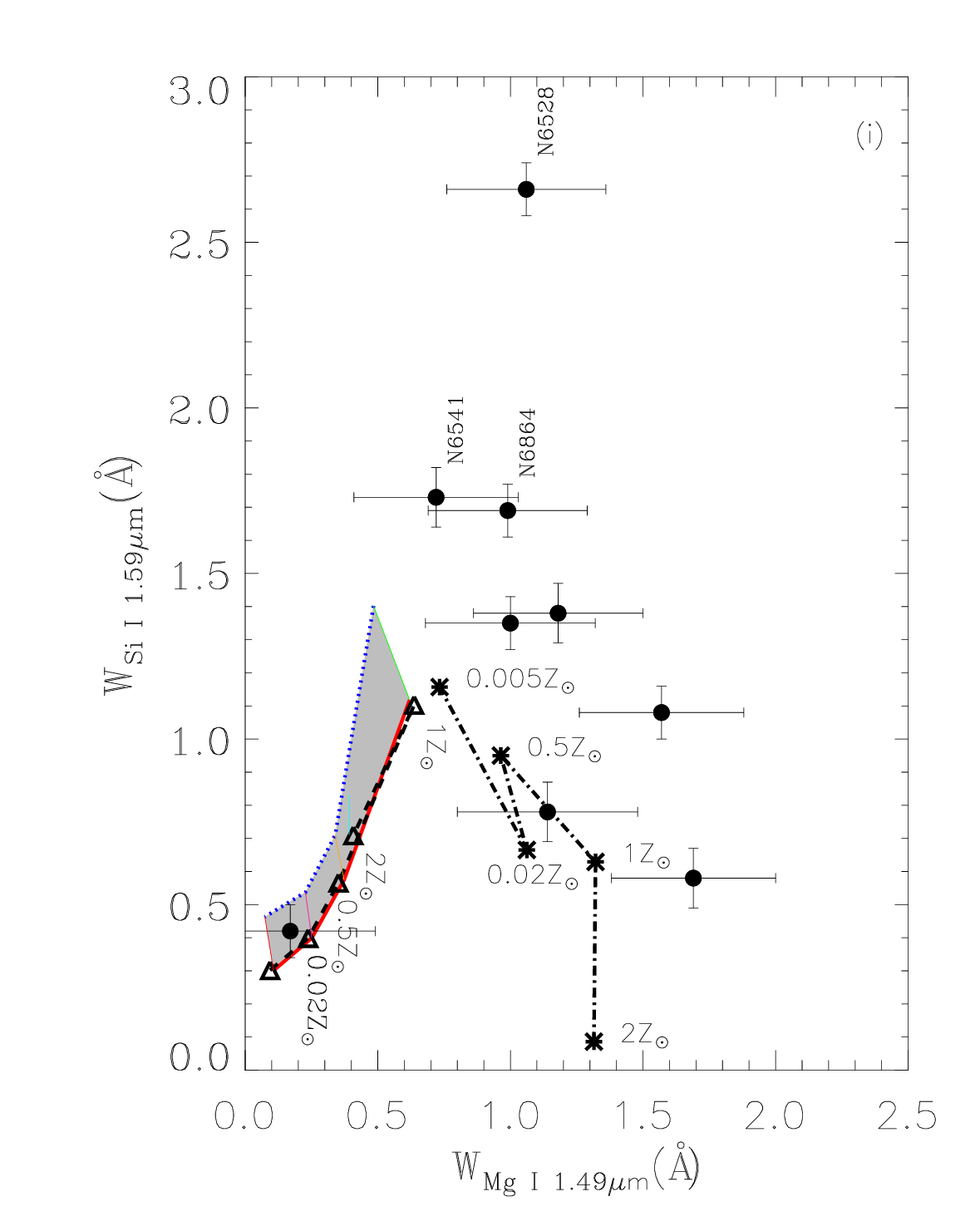} %i
\end{minipage}

\caption{Comparison of the measured \w\ with model predictions. The shaded region represents models predictions for
ages from 4 up to 15 Gyr (dotted and solid lines represent the
limits), with metallicities between $\frac{1}{200}\,Z\odot$ and 2\,$Z\odot$. Open triangles joined by 
a dashed line represent models which consider the blue horizontal branch. Asterisks joined by the dot-dashed 
line represent models with 1 Gyr old. Metallicities are labelled, except the 
lowest value, which we left out for display purposes.}
\label{mod_ew}
\end{figure*}

%%%%%%%%%%%%%%%%%%%%%%%%%%%%%%%%%%%%%%%%%%%%%%%%%%%

\begin{figure*}
\begin{minipage}[b]{0.33\linewidth}
\includegraphics[scale=0.30]{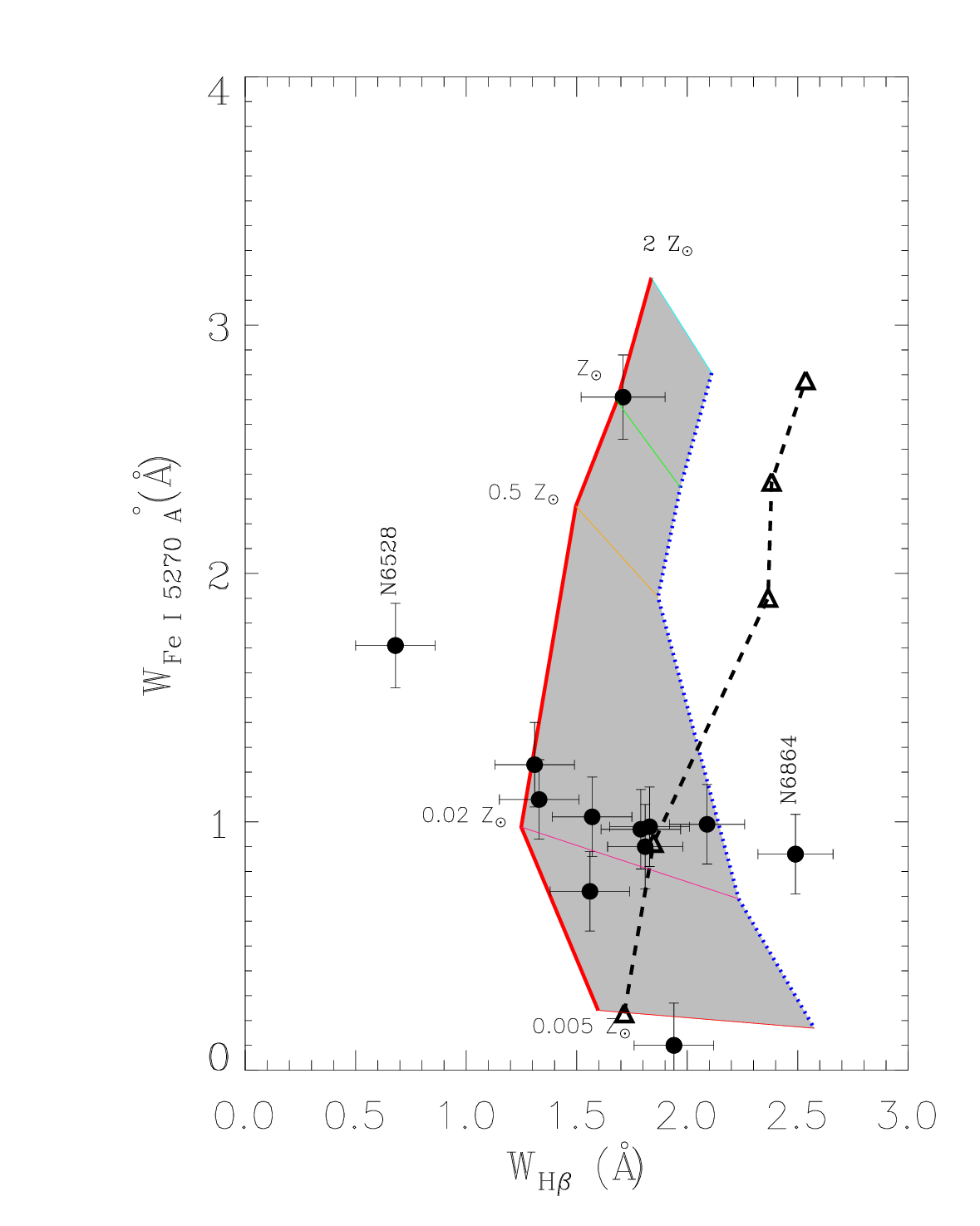} %a
\end{minipage}
\hfill
\begin{minipage}[b]{0.33\linewidth}
\includegraphics[scale=0.30]{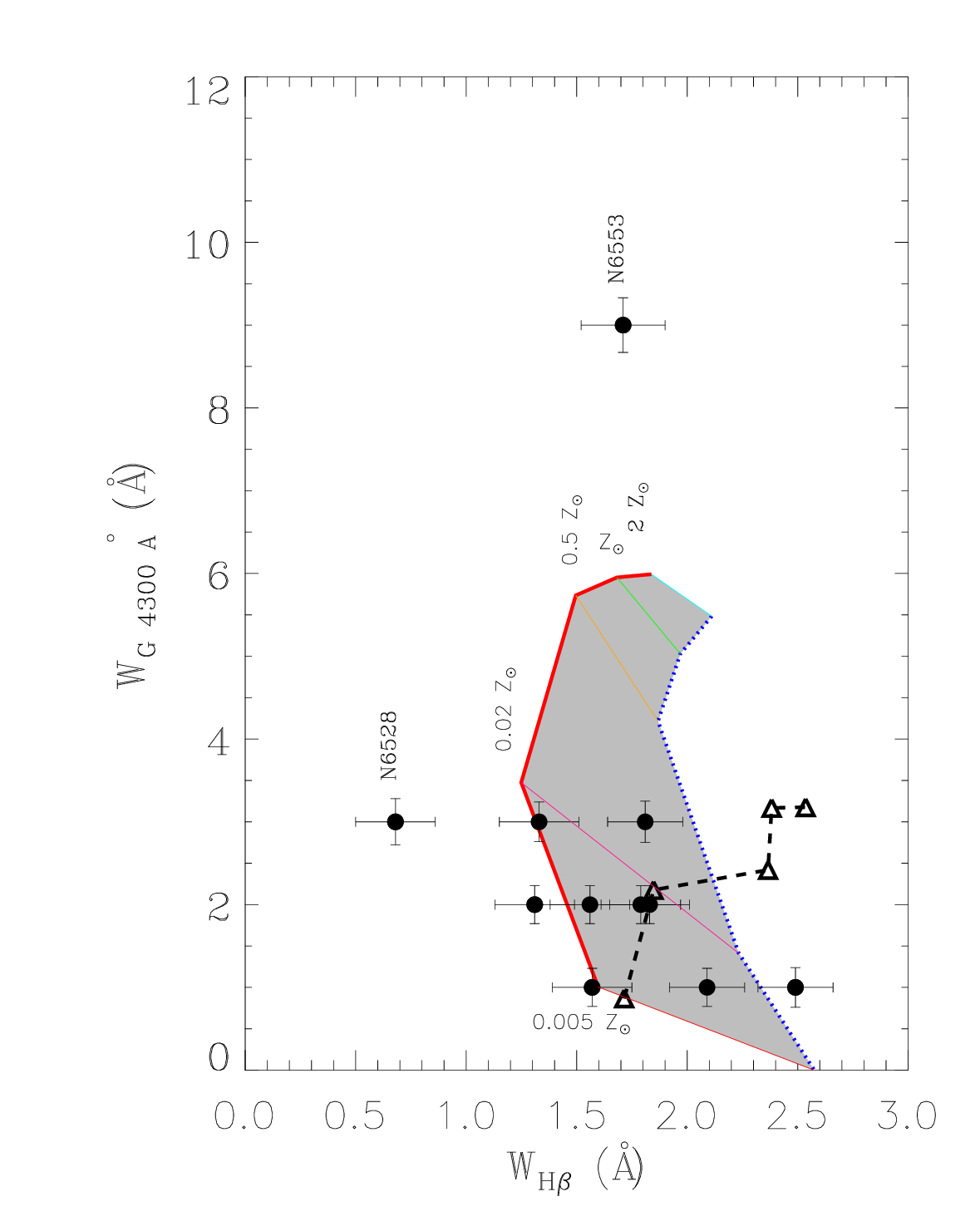} %b
\end{minipage}
\begin{minipage}[b]{0.33\linewidth}
\includegraphics[scale=0.30]{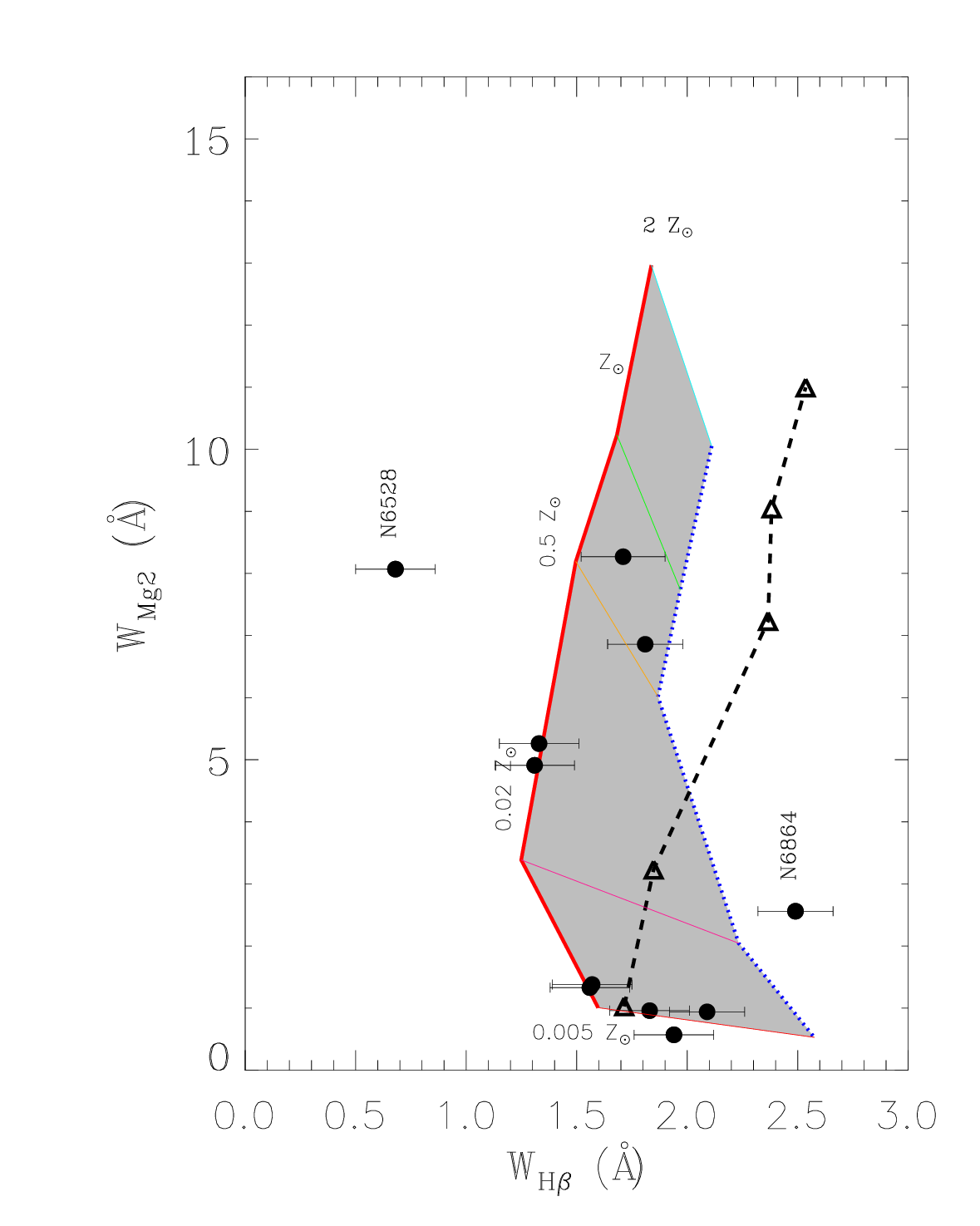} %c
\end{minipage}
\caption{Same as Fig.~\ref{mod_ew} but for optical \w\ and without the 1\,Gyr models.}
\label{mod_ewopt}
\end{figure*}

%%%%%%%%%%%%%%%%%%%%%%%%%%%%%%%%%%%%%%%%%%%%%%%%%%%

\begin{figure*}
\begin{minipage}[b]{0.33\linewidth}
\includegraphics[scale=0.30]{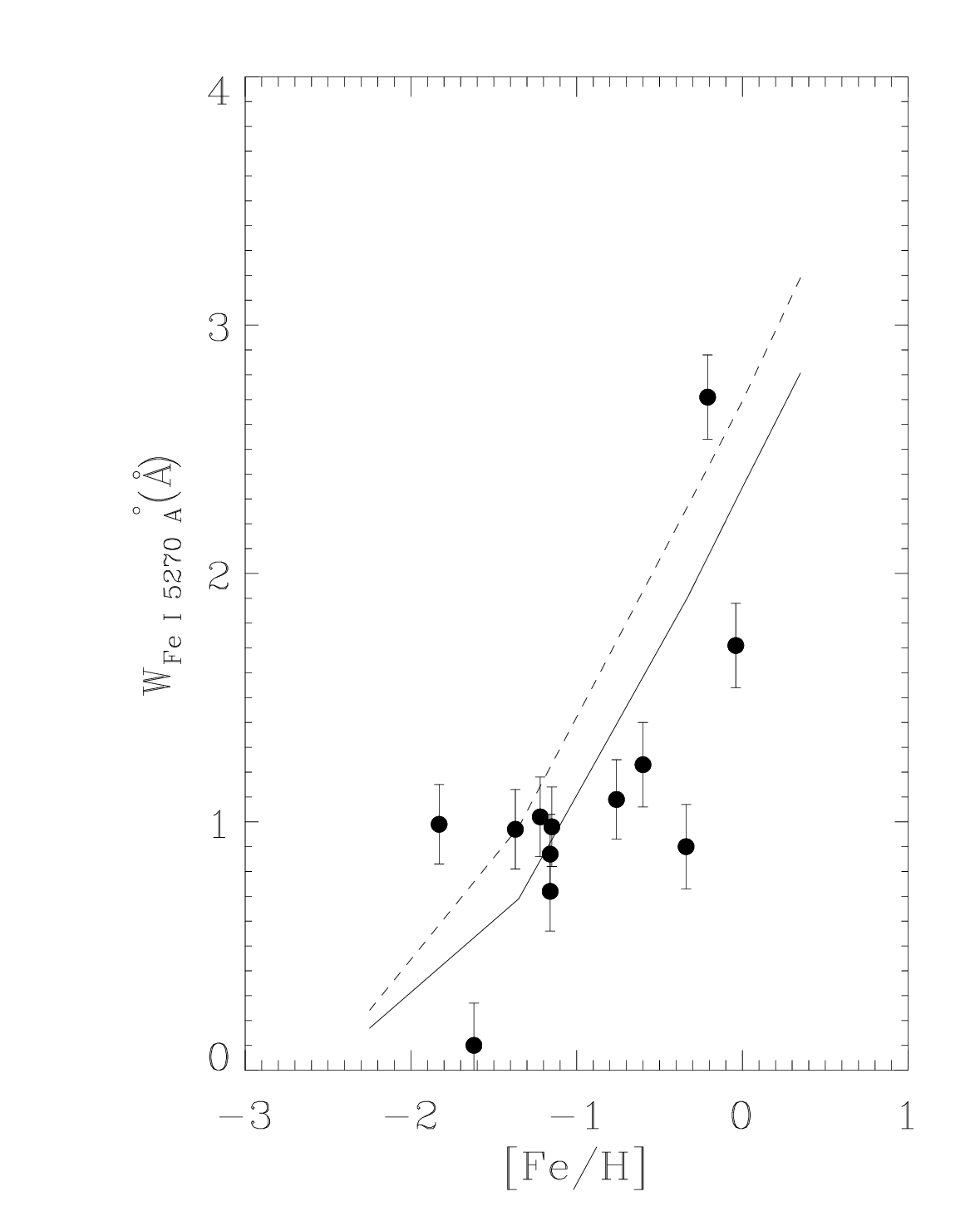} %a
\end{minipage}
\hfill
\begin{minipage}[b]{0.33\linewidth}
\includegraphics[scale=0.30]{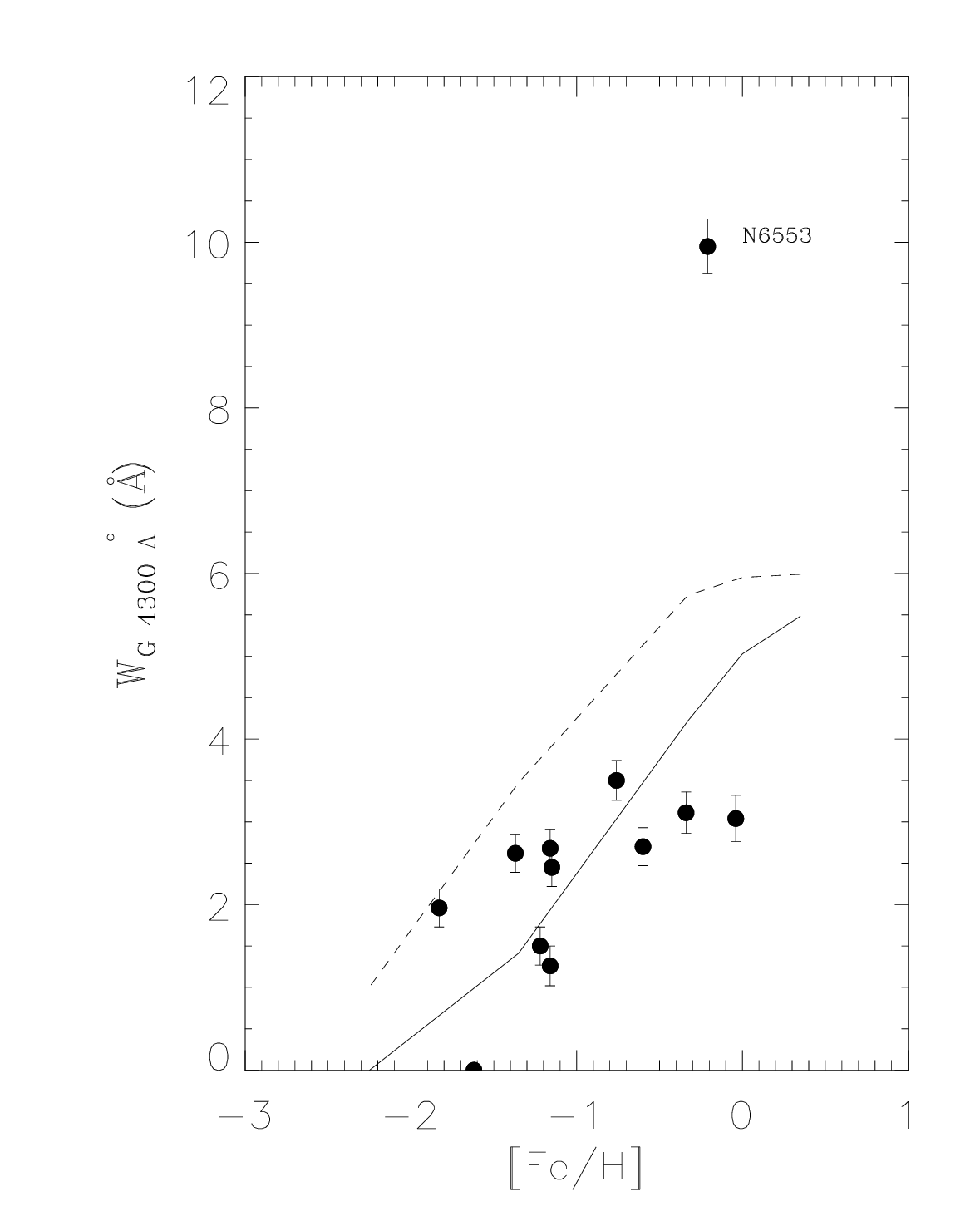} %b
\end{minipage}
\begin{minipage}[b]{0.33\linewidth}
\includegraphics[scale=0.30]{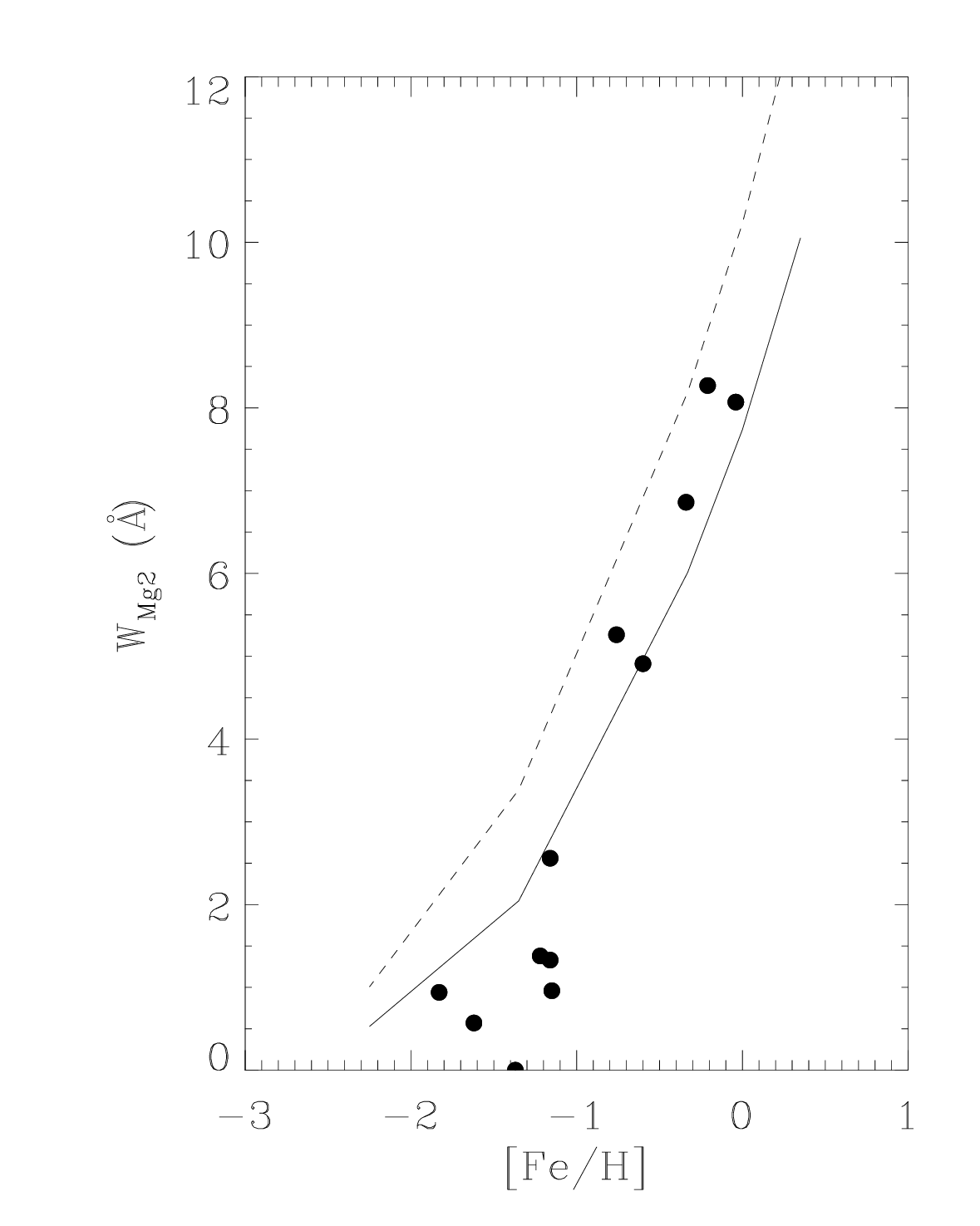} %c
\end{minipage}
\caption{Dependence of selected optical indices on [Fe/H]. Full and dashed lines are 4\,Gyr and 13\,Gyr old models, respectively.}
\label{index_metal}
\end{figure*}

%%%%%%%%%%%%%%%%%%%%%%%%%% Equivalent Width %%%%%%%%%%%%%%%%%%%%%%%%%%

Indices, such as \w\ of spectral absorption lines, can be seen as a compressed, but highly informative, 
representation of the whole spectrum \citep{cid07}. Thus, they are one of the more suitable ways to 
compare the observables with model predictions.  To compare observations with models, NIR spectral indices 
(this work), were plotted against each other and versus optical \w. Some of the most significant correlations
are shown in Fig.~\ref{mod_ew}, together with model predictions.

In Fig.~\ref{mod_ew}a we compare the relation between NIR \ion{Fe}{i} 1.58\,\mc\ and
the optical \ion{Fe}{i} 4531\,\AA\ with model predictions. The shaded region represents the model values 
for ages from 4 up to 15 Gyr (dotted and solid lines represent the
limits), with metallicities between $\frac{1}{200}\,Z_{\odot}$ and 2\,$Z_{\odot}$. Clearly, the models
can reproduce the \w\ of the optical \ion{Fe}{i} absorption line of the star clusters (i.e. 
old ages and $Z\leq Z_{\odot}$). However, they only predict the values of the NIR \ion{Fe}{i} line in
about half of our sample.  Similar conclusions are obtained 
when considering Figs.~\ref{mod_ew}b and \ref{mod_ew}c, in the latter case, however, for the NIR \ion{Si}{i} line.

\citet{cid10} have shown the importance of modelling the horizontal branch to properly
estimate the origin of blue stellar populations in integrated light analyses.
As M05 models consider the blue horizontal branch, we 
include such models in Fig.~\ref{mod_ew}a (open triangles joined by a dashed line). Note that we have only included
the 15\,Gyr old population because the 4\,Gyr models are not available for all metallicities. No significant
differences are observed in the prediction of blue horizontal branch and standard models in the case of  the \ion{Fe}{i}
and \ion{Si}{i} lines (Figs.~\ref{mod_ew}a to \ref{mod_ew}c).

Similarly to Fig.~\ref{mod_ew}a, we compare the optical G band and the 2.29\,\mc\ CO band 
in Fig.~\ref{mod_ew}d. As in the case of the \ion{Fe}{i} lines, the observed optical band values are almost 
all in agreement with the model predictions. However, in the case of the 2.29\mc\ CO band, the models
predict lower values than the observations for $\sim$ 50\% of the sample. 
A similar trend occurs for  H$\beta$ (Fig.~\ref{mod_ew}e)  and \ion{Na}{i} 5895\AA\ (Fig.~\ref{mod_ew}f) .

Regarding the presence of blue horizontal branch stars in the
models, they do describe better the optical absorption, i.e. they
predict lower values for the G band.  As expected, they also play an important role in determining the \w\ of H$\beta$. In contrast, the blue
horizontal branch models follow the standard ones concerning the \ion{Na}{i} line (Fig.~\ref{mod_ew}f).

The NIR \ion{Mg}{i} line and the optical Mg$\rm _2$  are compared with the models in Fig.~\ref{mod_ew}g.  
As for the Iron lines, the optical feature is properly predicted by the models, while they fail in the case 
of the NIR \ion{Mg}{i} line, since in this case the predicted values are about one order 
of magnitude lower than the observations. The blue horizontal branch models follow the standard ones.

In Fig.~\ref{mod_ew}h and \ref{mod_ew}i we compare NIR absorption lines against each other. It is clear that 
the models do not reproduce the observed strengths for \ion{Mg}{i} 1.49\mc. However, as discussed above
they do describe the measured values for \ion{Fe}{i} 1.58\mc\ and \ion{Si}{i} 1.59\mc\ in
about half of the sample. As expected, the blue horizontal branch models  are very similar 
to the standard ones. 

We also check the M05 models in the optical by comparing the \w\ of optical absorption lines against each other in Fig.~\ref{mod_ewopt}.
Clearly, the models reproduce the absorption line strengths measured in almost all objects.  
Although a dependence on metallicity is already apparent in Figs.~\ref{mod_ewopt}-\ref{mod_ew}, we show this more 
explicitly Fig.~\ref{index_metal}, for selected optical indices. Both Fe{\sc i} 5270\AA\ and Mg$\rm _2$ present a mild correlation with [Fe/H] that,
to a lesser degree, also applies to the G-band.  The M05 models are sensitive to metallicity,
and they do reproduce the observations.
It is worth mentioning that there are two outliers in Fig.~\ref{mod_ewopt}, NGC\,6528 and NGC\,6864. The former is located close to the Galactic centre and,
thus, W$\rm _{H\beta}$ may be attenuated by reddening. 
Regarding NGC\,6864, there is a significant contribution of blue horizontal branch 
stars \citep[see Fig. 5 of][]{kravtsov07}, which may be enhancing the H$\beta$ absorption.

The perception gained when considering absorption features is that M05 
models, which up to date are the most suitable to describe almost all absorption lines/bands observed in the NIR, 
do properly reproduce the optical absorption line strengths. Regarding the NIR absorption lines, 
the models do underestimate the strengths of \ion{Mg}{i} 1.49\mc, but they can properly reproduce the 
observed \w\ of  \ion{Fe}{i} 1.58\mc, \ion{Si}{i} 1.59\mc, and CO 2.29\mc, in about half of our sample.

In the case of CO 2.29\mc\ \w, similar results were obtained by \citet[][see their Fig.~9]{lyubenova10} for old and 
metal-poor Large Magellanic Cloud (LMC) globular clusters. These authors also show that a large 
fraction of carbon stars would mimic the spectrum of a younger cluster. To check this 
hypothesis, we overplot in Fig.~\ref{mod_ew} 1\,Gyr models
(asterisks joined by dot-dashed line). Interestingly, the intermediate age models can describe the 
observed $W_{\rm CO\,2.29\mu m}$ in about half of our sample. Thus, we suggest that the NIR light in the core
of these clusters (NGC\,104, NGC\,362, NGC\,2808, NGC\,6388, NGC\,6528, NGC\,6553) may, perhaps, be 
dominated by C-rich stars.  Note that, the metallicity distribution of the 
clusters is heterogeneous (see Tab.~\ref{prop}), thus these large values of \w\ of the CO 2.29\mc\ are probably not related 
to metallicity. This suggests a possible carbon-fraction/age degeneracy that affects NIR 
stellar populations. Indeed, according to \citet{marigo09} (see their Fig. 5), the $K$
luminosity fraction of TP-AGB stars in $10$\,Gyr old SSPs of different metallicities has been shown to grow from about 15\%
(at [Z/Z$_{\odot}$]=0.0) to 50\% (at [Z/Z$_{\odot}$]=-2.3).  We remind that the 1\,Gyr models in Fig.~\ref{mod_ew} are not related 
to cluster age. Instead, thei simply reflect the presence of C-rich stars in some globular clusters. Also, it is interesting to remark
that C-rich stars are only important in the NIR, beeing essentially unnoticed in the optical  (see Fig.~\ref{mod_ewopt}).

Interestingly, the \ion{Fe}{i} 1.58 \mc\ atomic lines are well reproduced by the 1\,Gyr population in 
about half of the sample (NGC\,362, NGC\,6388, NGC\,6528, NGC\,6553 and NGC\,6864). In the case of 
\ion{Mg}{i} 1.49\mc, it occurs for almost all objects (NGC\,104, NGC\,2808, NGC\,6388, NGC\,6528 and NGC\,6864).
This suggests that the inclusion of empirical spectra of C- and O-rich stars in the models may also
have an important effect in the strength of metallic lines.  Another possibility lies in the fact 
that we are studying $\alpha$-enhanced Galactic globular clusters and, as the models do not include
$\alpha$-enhancement, this might explain why the models cannot reproduce the \w\  of absorptions involving
$\alpha$-elements.

%\subsubsection{Color - color diagrams}

Another powerful way to test SSPs models is by comparing observed and predicted colours. 
For this purpose, we collected the optical to NIR colours of our star clusters from the literature. These values 
are listed in Tab.~\ref{prop}, together with some basic properties of the cluster sample.

\begin{figure*}
\begin{minipage}[b]{0.45\linewidth}
\includegraphics[scale=0.45]{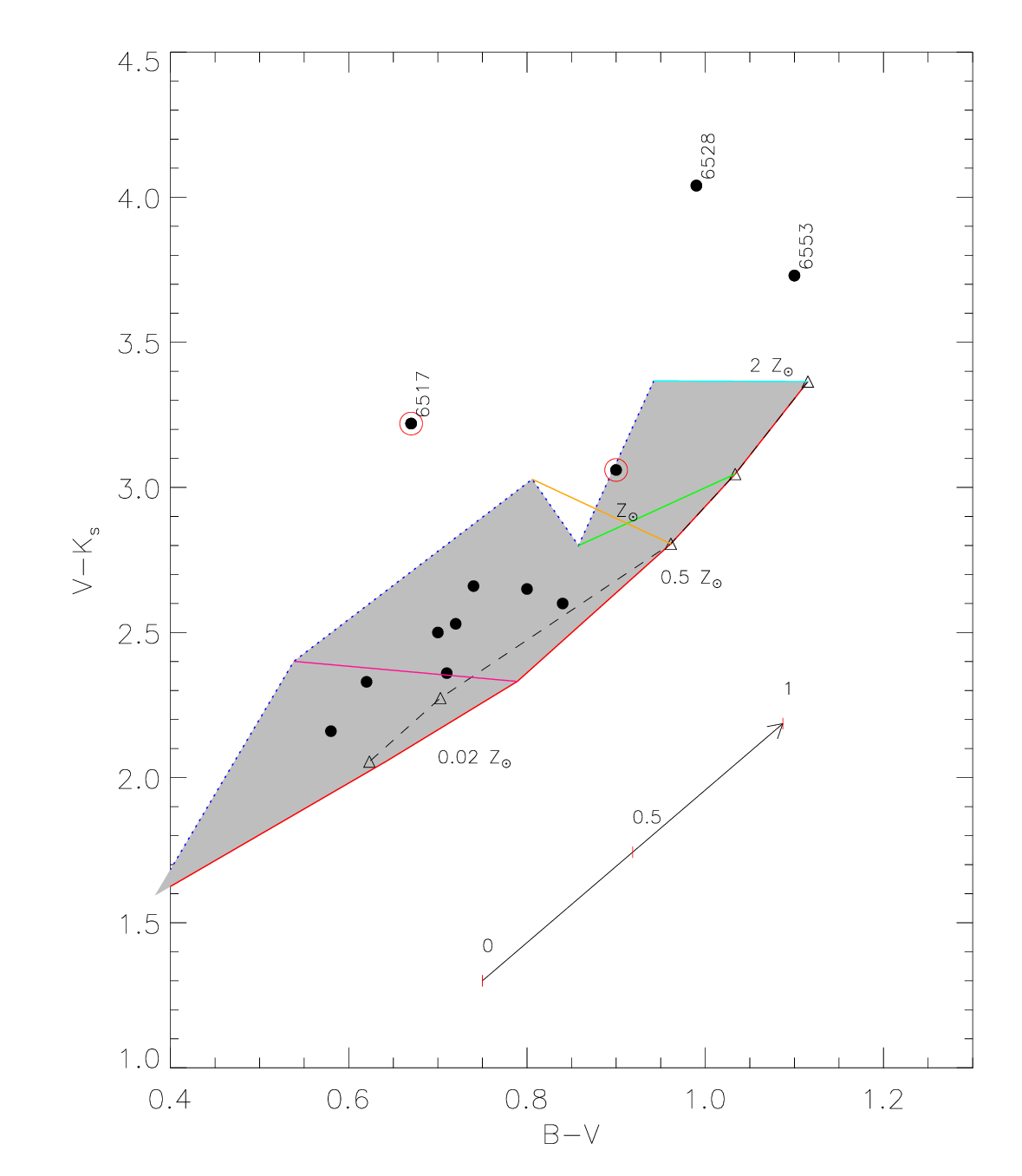}
\end{minipage}
\hfill
\begin{minipage}[b]{0.45\linewidth}
\includegraphics[scale=0.45]{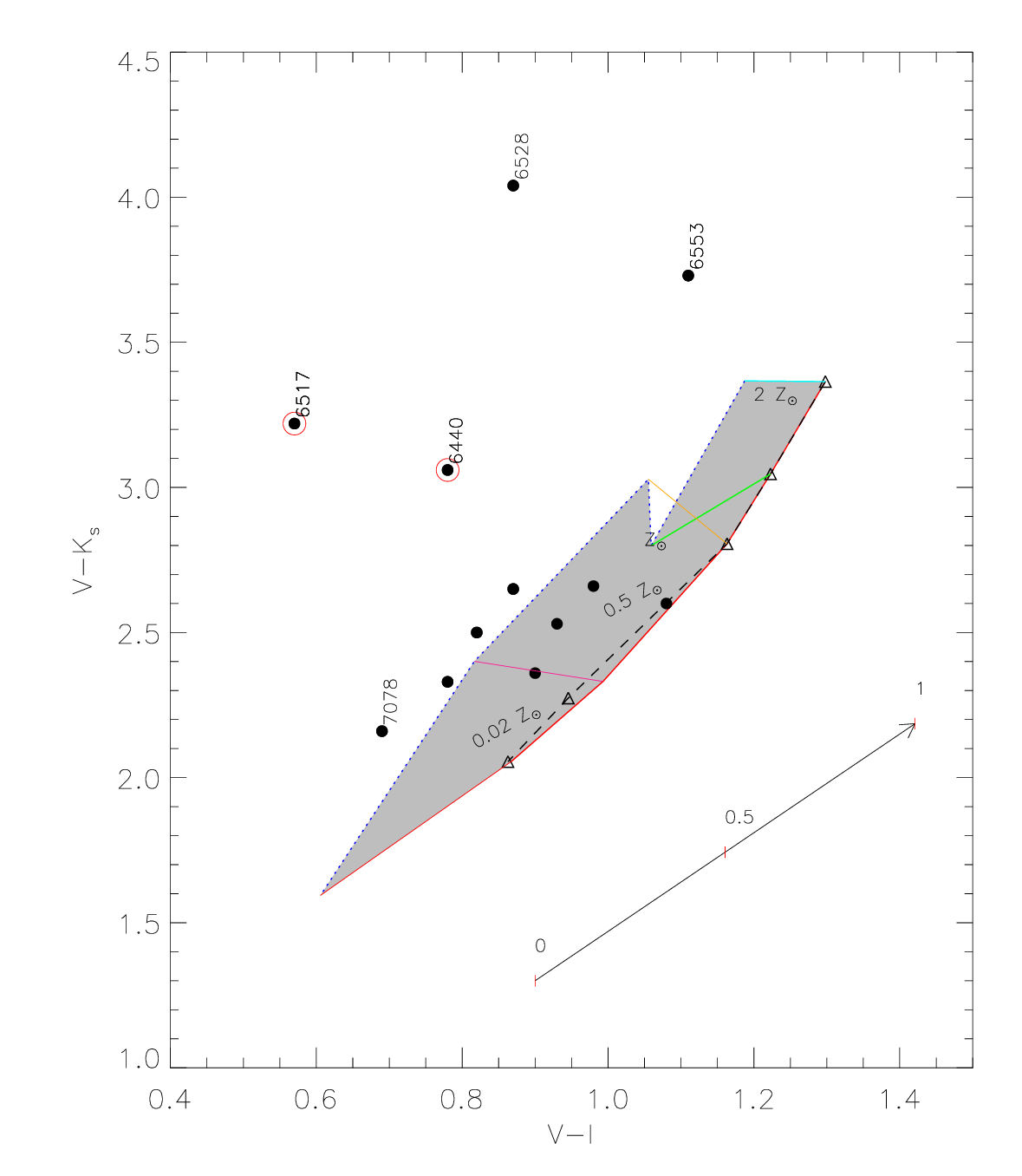}
\end{minipage}
\caption{Colour-Colour diagrams. Open circles indicate objects with E(B-V)$>$1.0. A reddening
vector is shown for 0$\leq A_{\rm v} \leq$1. The lines are the same as in Fig~\ref{mod_ew}, 
 but the young age is 2\,Gyr.}
\label{cmd}
\end{figure*}

We compare the observed colours with the models prediction in Fig.~\ref{cmd}. Note that the optical 
colours were reddening-corrected using the \citet{ccm89} extinction law, and the E(B-V) values 
from \citet{bica06}. Clearly if standard errors,  in colours, 
are taken into account, the models are able to describe almost all the observations.  However, 
there are some outliers (NGC\,6440, NGC\,6517, NGC\,6528 and NGC\,6553) for which it is not 
possible to determine whether or not the models apply,  or if the photometric uncertainties are large, since these 
objects are located very close to the Galactic centre, and thus are more likely contaminated by field stars.
Note that these objects are located in the Galaxy bulge (see Tab.~\ref{prop}), 
and therefore, precise measurements are difficult.  In addition, uncertainties in SSPs due to
stellar tracks and EPS codes in the $V-K$ colour are typically 0.25 mag \citep[see][M05]{charlot96}.
In this way, we suggest that M05 models, in general, are able to predict the observed 
color values for the optical.   For the NIR, however, they do  systematically predict bluer 
colours, and a 2\,Gyr SSP is required to properly reproduce the observed NIR colours. Thus, they may underestimate
the age for the old population.

The above results lead us to conclude that M05 models can provide 
reliable information on the NIR stellar population  of galaxies only when \w\ and colours are 
taken together, in other words, \w\ and continuum fluxes should be simultaneously fitted \citep[e.g][]{riffel08}, 
alternatively  the whole underlying spectrum should be used \citep[e.g.][]{riffel09,rogemar10,martins10}. 
However, the results should be taken with caution, since the models tend to predict results 
biased towards young ages. A more robust test is fit the whole globular clusters spectra with models. However, this 
is out of the scope of this paper and we leave it for a forthcoming publication 
(Ruschel-Dutra et al., 2010 {\it in preparation}).

%%%%%%%%%%%%%%%%%%%%%%%%%% Color - Color %%%%%%%%%%%%%%%%%%%%%%%%%%

\section{Conclusions}\label{conclusion}

 We use NIR SOAR/OSIRIS integrated spectra of 12 Galactic globular clusters 
to test M05 NIR EPS models, and to provide spectral observational 
constraints to calibrate future NIR EPS models. M05 models are used for 
being, up to date, the most suitable  to describe absorption lines/bands in the NIR. Our main conclusions are:

\begin{itemize}

\item The spectra of the cluster sample appear qualitatively 
similar in most of the NIR absorption features. 

\item Many atomic absorption features like: $\lambda$ 1.49\mc\ \ion{Mg}{i}, $\lambda$ 1.58\mc\ Fe {\sc i}/\ion{Mg}{i}, $\lambda$ 1.59\mc\ 
Si {\sc i}, $\lambda$ 1.71\mc\ Mg {\sc i}, $\lambda$ 2.21\mc\ Na {\sc i} 
and  $\lambda$ 2.26\mc\ \ion{Ca}{i} as well as the $\lambda$ 1.62\mc, $\lambda$  2.29\mc\ CO and 
$\lambda$ 2.05\mc\ CN molecular bands are clearly detected and identified in the spectra. The \w\ of these features were
measured, as well as the optical \w\ of of G band (4300\AA),
MgH (5102\AA), and FeI (4531\AA). The
globular clusters observations (\w\ and colours) were compared with models predictions with ages from 4 up to 15 Gyr,
with metallicities between $\frac{1}{200}\,Z\odot$ and 2\,$Z\odot$.

\item  M05 models are able to properly reproduce the
   optical  \w\ observed in globular clusters.

\item The models do underestimate the strength of \ion{Mg}{i} 1.49\mc, but they can
properly reproduce the observed \w\ of  \ion{Fe}{i} 1.58\mc, \ion{Si}{i} 1.59\mc\ and CO 2.9\mc\  
in about half of our sample. For the remaining objects, we needed to consider intermediate-age 
populations. Thus, we suggest that the presence of C- and O-rich stars in the models is important 
to reproduce the observed strengths of metallic lines. Another possibility is the lack of 
$\alpha$-elements enhancement in the models.

\item    No significant differences are observed in the prediction of blue horizontal branch and standard 
models in the case of the \ion{Fe}{i} lines (optical and NIR). A similar trend is observed for the \ion{Na}{i} 5895\AA. While in 
the case of the G-band the models which include blue horizontal branch do describe better the observations. No differences between blue horizontal 
branch and standard models are found for the NIR observations.

\item   In general, M05 models can reproduce the observed 
colours in the optical, while in the NIR they tend to underestimate the age of the old population.

\end{itemize}

The NIR spectral region is the most convenient one to study the stellar content of highly obscured sources. Thus, well 
calibrated and accurate NIR models is fundamental. Besides, observations using 
adaptive optics, to properly correct for atmosphere effects, have become commonly used
in the NIR region, thereby allowing for high quality spectroscopy \citep[e.g.][]{rogemar10}. 
The availability of detailed, high spectral resolution, and well calibrated stellar population models 
in the NIR will open a new window to disentangle the NIR 
stellar content in the inner few pc of nearby galaxies. Thus, the data presented in this paper 
will be important for the new generation of NIR EPS models.

\section*{Acknowledgements}
We thank an  anonymous referee for interesting comments. R.R. thanks for the 
support from the Brazilian funding agency Capes and CNPq. The authors are 
grateful to C. Maraston for useful discussions.
JFCSJ thanks Brazilian agency FAPEMIG (grant APQ00117/08). ARA thanks to CNPq for
partial support through grant 308877/2009-8.
This work has been done with observations
from the SOAR telescope, a collaboration among The Minst\'erio da Ci\^encia e
Tecnologia/Brazil, NOAO, UNC and MSU.

\end{document}